\documentclass[%
 reprint,
 amsmath,amssymb,
 aps,
prl
]{revtex4-2}

\usepackage{graphicx}
\usepackage{dcolumn}
\usepackage{bm}
\usepackage{braket}
\usepackage{bibunits}
\usepackage{placeins}
\usepackage{stackengine}

\usepackage{amsmath}
\usepackage{tabstackengine}
\stackMath
\usepackage{float}

\usepackage{hyperref}
\hypersetup{colorlinks,allcolors=magenta}

\def\mathclap#1{\text{\hbox to 0pt{\hss$\mathsurround=0pt#1$\hss}}}

\begin{document}

\preprint{APS/123-QED}

\title{Strongly Enhanced Electronic Bandstructure Renormalization by Light \\ in Nanoscale Strained Regions of Monolayer MoS$_2$/Au(111) Heterostructures}%

\author{Akiyoshi Park$^{1,2}$}
\author{Rohit Kantipudi$^{1}$}
\author{Jonas G{\"o}ser$^{1,3}$, \\ Yinan Chen$^{1,2}$, Duxing Hao$^{1,2}$, and Nai-Chang Yeh$^{1,2}$}
\thanks{Corresponding Author: \href{mailto:ncyeh@caltech.edu}{ncyeh@caltech.edu} }
\affiliation{$^1$ Department of Physics, California Institute of Technology, Pasadena, CA 91125, USA}
\affiliation{$^2$ Institute for Quantum Information and Matter, California Institute of Technology, Pasadena, CA 91125, USA}
\affiliation{$^3$ Fakultät für Physik, Munich Quantum Center (MQC), and Center for NanoScience (CeNS), Ludwig-Maximilians-Universität München, Geschwister-Scholl-Platz 1, 80539 München, Germany}

\date{\today}

\begin{abstract}
Understanding and controlling the photoexcited quasiparticle (QP) dynamics in
monolayer transition metal dichalcogenides (TMDs) lays the foundation for exploring the strongly interacting, non-equilibrium two-dimensional (2D) quasiparticle and polaritonic states in these quantum materials and for harnessing the properties emerging from these states for optoelectronic applications. In this study, scanning tunneling microscopy / spectroscopy
(STM/STS) with light illumination at the tunneling junction is performed to investigate the QP dynamics in monolayer MoS$_2$ on an Au(111) substrate with nanoscale corrugations. The corrugations on the surface of the substrate induce nanoscale local strain in the overlaying monolayer MoS$_2$ single crystal, which result in energetically favorable spatial regions where photoexcited QPs, including excitons, trions, and electron-hole plasmas, accumulate. These strained regions exhibit pronounced electronic bandstructure renormalization as a function of the photoexcitation wavelength and intensity as well as the strain gradient, implying strong interplay among nanoscale structures, strain, and photoexcited QPs. In conjunction with the experimental work, we construct a theoretical framework that integrates non-uniform nanoscale strain into the electronic bandstructure of a monolayer MoS$_2$ lattice using a tight-binding
approach combined with first-principle calculations. This methodology enables better understanding of the experimental observation of photoexcited QP localization in the nanoscale strain-modulated electronic bandstructure landscape. Our findings illustrate the feasibility of utilizing nanoscale architectures and optical excitations to manipulate the local electronic
bandstructure of monolayer TMDs and to enhance the many-body interactions of excitons, which is promising for the development of nanoscale energy-adjustable optoelectronic and photonic technologies, including quantum emitters and solid-state quantum simulators for interacting exciton polaritons based on engineered periodic nanoscale trapping potentials.
\end{abstract}

\begin{bibunit}

\maketitle
\section{Introduction}
The reduced Coulomb screening in two-dimensional (2D) semiconducting transition metal dichalcogenides (TMDs) prompts strong exciton binding energies ($E_\mathrm{b}$) that are significantly larger than those in three-dimensional semiconductors \cite{PhysRevB.86.241201, PhysRevLett.113.076802, PhysRevLett.113.026803, Feng2012, PhysRevLett.111.216805}. The weak dielectric screening of excitons in 2D-TMDs results in substantial many-body correlations and exotic excited phases that are not existent in the equilibrium states of traditional semiconductors. In particular, excitonic interactions have been reported to generate substantial renormalization effects on the ground state electronic bandstructure \cite{Ugeda2014, Cunningham2017, Pogna2016, Sie2017, Chen2022, Chernikov2015, PhysRevLett.122.246803}, with the degree of bandstructure modifications amplified to alter the bandgap up to 500 meV by increasing the exciton density \cite{Chernikov2015, PhysRevB.106.L081117}. Hence, photoexcited QP effects, such as excitonic interactions and photo-induced free carrier screening, provide an adjustable control for the optoelectronic properties of TMDs, which is a valuable feature for various device applications.
\begin{figure*}[t]
\includegraphics[width = 0.8\textwidth]{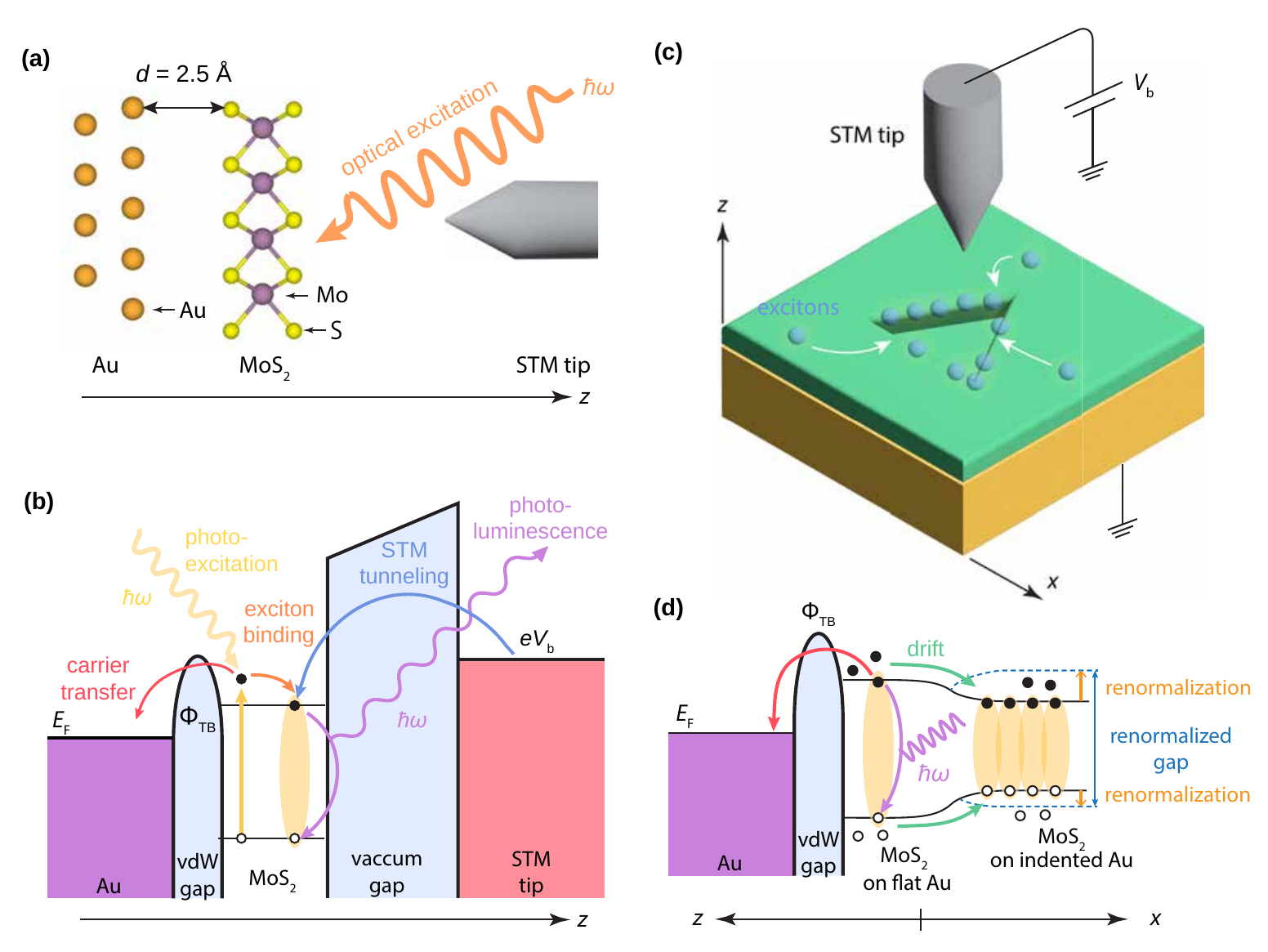}
  \caption {Charge carrier dynamics: (a) A schematic diagram of the experimental set-up of STM studies on an optically excited ML-MoS$_2$ on Au(111) heterostructure. (b) The corresponding energy diagram of the STM tunneling junction showing the various possible pathways of charged carrier transfer processes, including interlayer tunneling from ML-MoS$_2$ through the vdW gap to Au and photoluminescence via intralayer electron-hole recombination. (c) A three-dimensional (3D) schematic diagram indicating the aggregation of photoexcited QPs to regions of ML-MoS$_2$ strained by a nanoscale indentation in the Au substrate. (d) An energy diagram illustrating a competing pathway of in-plane carrier drift to a strained region in ML-MoS$_2$ with a modulated band energy, versus interlayer carrier tunneling through the vdW barrier into the Au substrate and native radiative recombination.}
  \label{Fig1}
\end{figure*}

However, there exists ambivalence to the substantial many-body interactions of photoexcited QPs in the TMDs. While it causes the emergence of exotic steady-state phenomena, it becomes a challenge to collect the photo-induced electron/hole free carriers \cite{Massicotte2018}, which is an important facet for optoelectronic applications in photovoltaics and photodetectors that hinge on efficient collection of photo-induced carriers into metal electrodes \cite{GrubišićČabo2015, Krane2016, Datye2022}. Hence, for the sake of application of TMDs in tunable optoelectronics, both efficient formation of excitons to modify the bandstructure and an efficient method of carrier extraction to improve the performance of optoelectronic conversion is necessary.

An efficient extraction of electrons and holes from excitons can be achieved through interfacing the TMD with a metallic surface such as Au(111) \cite{GrubišićČabo2015, Goswami2019}. Interlayer transfer of charge carriers across the MoS$_2$ and Au interface is dictated by carrier tunneling through a barrier constituted by a van der Waal (vdW) gap. The interaction between MoS$_2$ and Au has been both theoretically \cite{Velický2018, Sarkar2021} and experimentally shown to be dominated by vdW interactions, owing to lack of significant orbital overlap between Au and MoS$_2$ \cite{PhysRevX.4.031005, Tumino2020}. The width of this vdW barrier is determined by the separation between the Au atom and the interfacial S atom, ranging from 2.51 to 2.79 \AA$\hspace{1 pt}$ (Fig. \ref{Fig1} (a)) \cite{Sarkar2021, Tumino2020, PhysRevLett.108.156802}, which is substantially greater than the 2.156 \AA$\hspace{1 pt}$ bond length of covalently interacting Au-S \cite{Kokkin2015}. Moreover, the substantial junction barrier height of $\Phi_\mathrm{TB}$ = 0.67 - 0.92 eV \cite{PhysRevX.4.031005} further support the notion that charge injection/extraction across the vdW gap is primarily carried out through a tunneling process.

Given that MoS$_2$ becomes electron-doped as a result of interfacing with Au, where the Fermi level ($E_\mathrm{F}$) is pinned right below the conduction band minima \cite{PhysRevX.4.031005, PhysRevB.93.165422}, tunneling of excited electrons from the conduction band of MoS$_2$ to Au(111) is a favorable photoexcited QP deactivation pathway in addition to native exciton recombination within MoS$_2$ (Fig. \ref{Fig1}(b)). In the case of excitons, although the binding energy of an exciton would resist the transfer of electrons into the Fermi sea of Au, due to the large static dielectric constant of Au, the exciton binding energy in MoS$_2$ adjacent to Au is substantially reduced to $E_\mathrm{b} \approx 90$ meV \cite{Park_2018}, thereby increasing the tendency of disassociating the excitons in monolayer (ML) MoS$_2$ into free charged carriers for tunneling into the underlying Au substrate. While it is known that photoluminescence (PL) spectroscopy exhibits a reduced PL yield from MoS$_2$ when interfaced with Au owing to charge transfer, nanoscale local measurements performed by STM-induced electroluminescence have demonstrated that MoS$_2$ on an Au substrate is still capable of hosting excitons and displaying excitonic physics including radiative decay and exciton-exciton annihilation \cite{D2RA05123K}. The charge transfer process for carriers tunneling from MoS$_2$ across the vdW gap to Au is relatively slow with a characteristic time of $\tau_\mathrm{CT} \approx 400 - 600$ fs \cite{Goswami2019, Xu2021} when compared with the exciton formation time of $\approx 30$ fs \cite{Trovatello2020}. This comparison of characteristic times signifies that exciton formation in our experiments, although competea with exciton purging due to the charge transfer process, excitons are constantly pumped into the system with a CW laser, thus allowing for consequential exciton population.  Moreover, given that the exciton diffusion coefficient is $\approx 20$ cm$^2$ s$^{-1}$ \cite{C3NR06863C}, within the time scale $\tau_\mathrm{CT}$ of the charge transfer process, excitons could travel a distance  $l \approx \sqrt{4D\cdot\tau_\mathrm{CT}} \approx 60$ nm in MoS$_2$ before they dissociated, allowing for magnified exciton-induced band renormalization at nanoscale strained regions.

In this study, we investigate the local electronic bandstructure renormalization effects of ML-MoS$_2$ on an Au(111) substrate through light-assisted scanning tunneling microscopy/spectroscopy (STM/STS). In general, we find that the electronic bandstructure renormalization in ML-MoS$_2$ due to photoexcited QPs is overpowered by the dominating effect of carrier transfer to the Au(111) substrate and the native QP recombination. However, this situation becomes substantially modified when ML-MoS$_2$ is strained by nanoscale corrugations in the Au(111) substrate. The strain on the ML-MoS$_2$ renormalizes the local electronic bandstructure and modulates the bandgap energies \cite{Scalise2012, PhysRevB.85.033305}, thus creating a funneling channel for photoexcited QPs to drift within the ML-MoS$_2$ \cite{doi:10.1126/sciadv.abm5236}, which competes with the processes of charge transfer to Au(111) and exciton recombination in ML-MoS$_2$ (Fig. \ref{Fig1}(c,d)) \cite{Feng2012}. In such an arrangement, the probability of the in-plane drift of photoexcited QPs to strained regions of band configurations with favorable energy is heightened relative to those of interlayer tunneling of photoexcited carriers through the vdW barrier into Au and the native intralayer QP recombination. Thus, regions of ML-MoS$_2$ with strain-induced band renormalization into which photoexcited QPs funnel will have more pronounced QP-induced band renormalization effects over less-strained regions. Here, we remark that the strain-induced structures utilized for carrier funneling in existing literature are typically on the scale of microns to hundreds of nanometers, as reported in studies such as Chaste \textit{et al.} (2018) and Li \textit{et al.} (2015) \cite{Chaste2018, Li2015}. In contrast, by employing strain features on a scale of merely several nanometers to angstroms, deeper potential wells for QP confinement may be achieved \cite{Chirolli_2019}, leading to higher densities of photoexcited QPs with substantially overlapped wavefunctions confined in the strain-induced potential wells, and therefore much enhanced light-induced renormalization effects on the quasiparticle LDOS. These findings thus hold promises for energy-adjustable quantum emitters \cite{YWang2021, Iff2019} and novel solid-state quantum simulators for exciton polaritons in periodic potential wells, offering valuable insights for advancements in photonic quantum information processing technologies \cite{O'Brien2009}.

\begin{figure}[t]
\includegraphics[width = 0.4\textwidth]{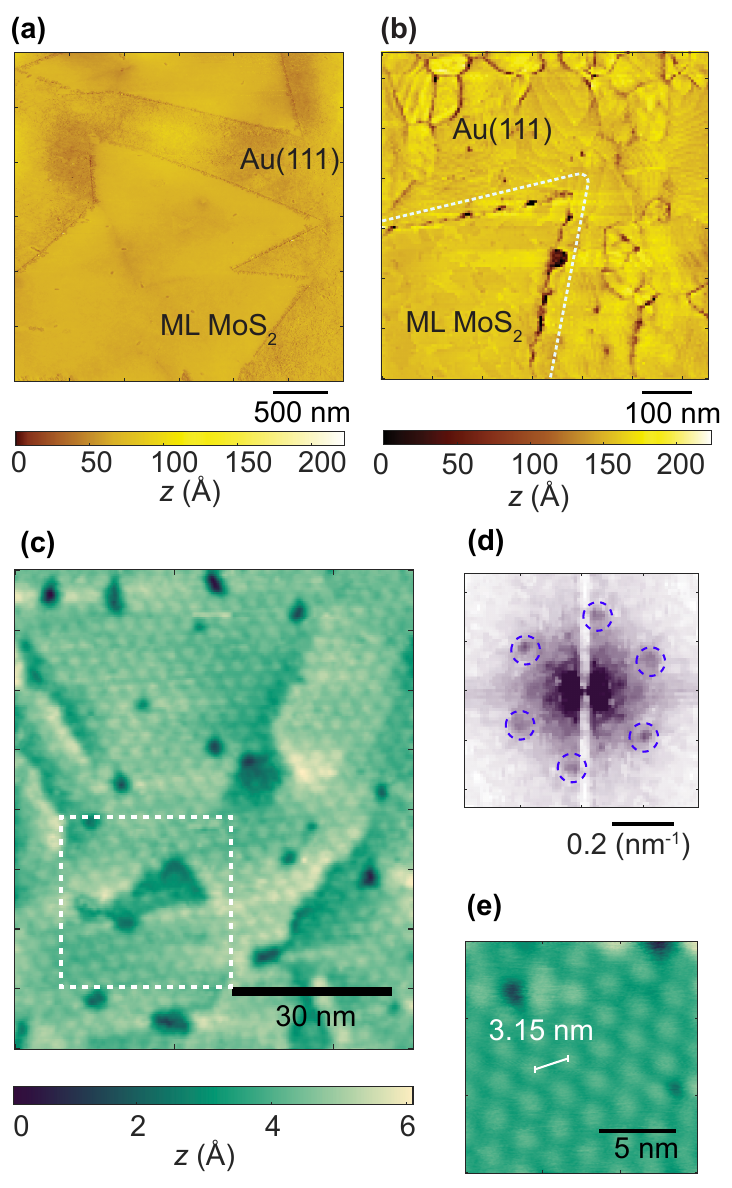}
  \caption {Topography of ML-MoS$_2$ on Au(111) heterostructure: STM topographic images of ML-MoS$_2$ on an Au(111) substrate, depicting (a) numerous triangular ML-MoS$_2$ single crystalline flakes of micrometer scale, and (b) a zoomed in image of one of the corners of the ML-MoS$_2$ flake on top of Au grains with terrace-like structures exhibiting its (111)-lattice plane surface. The dashed lines indicate borders between the ML-MoS$_2$ flakes and the Au substrate. (c) Nanoscale topographic images of ML-MoS$_2$ on an Au substrate. ML-MoS$_2$ is not perfectly flat, displaying wrinkles at the protrusion and indentations of the Au terraces underneath the MoS$_2$ monolayer. (d) A 2D fast Fourier transform of the topographic image of panel (c), representing the triangular Moir\'e superlattice. The hexagonal pattern in reciprocal space is indicated by the dashed circles. (e) A (15 nm × 15 nm) area of topographic image indicates that the Moir\'e superlattice periodicity is 3.15 nm.}
  \label{Fig2}
\end{figure}

\begin{figure}[t]
\includegraphics[width = 0.47\textwidth]{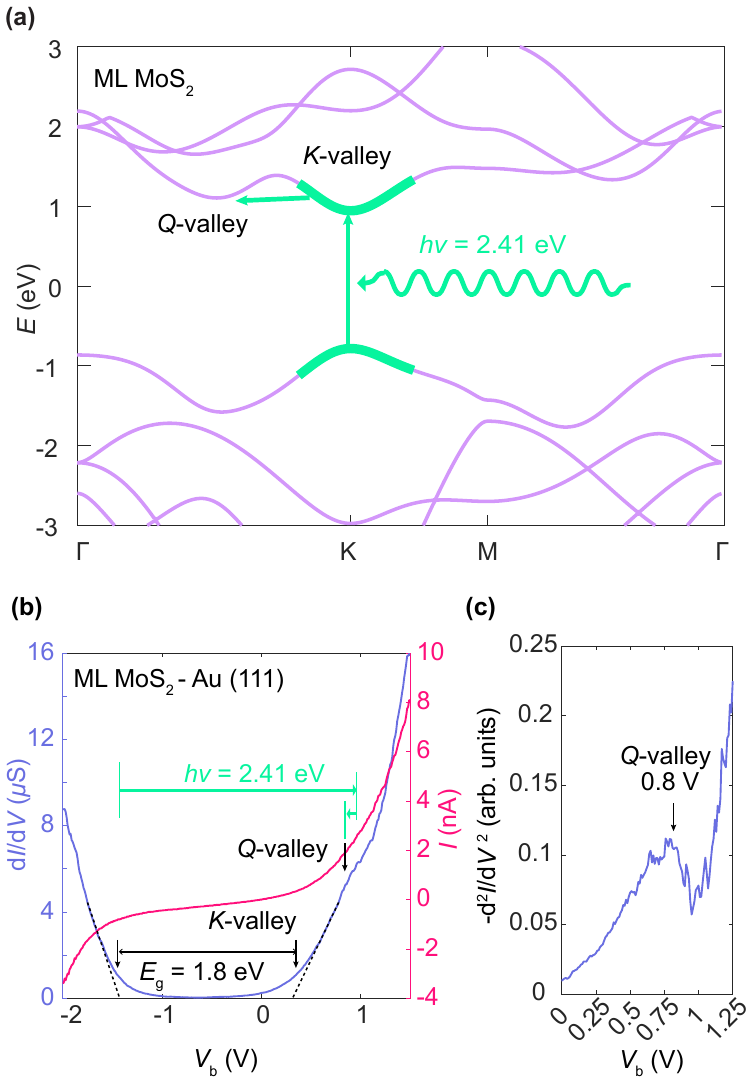}
  \caption {Electronic structure of the ML-MoS$_2$ on Au(111) heterostructure: (a) ML-MoS$_2$ bandstructure computed from the Maximally localized Wannier functions (MLWFs). The region of the bands highlighted in green indicate regions in the FBZ where $E_\mathrm{g}(k) \leq E_\mathrm{photon}$. Evidently, holes and electrons can populate both an energy of $\approx 0.5$ eV above and below the bandgap. (b) Representative $dI/dV$  and $I$ versus $V_\mathrm{b}$ spectra measured on a flat surface of MoS$_2$-Au(111) heterostructure. The spectrum was measured by fixing the STM tip in position at a set point of $V_\mathrm{b}=1$ V, $I_\mathrm{t}$ = 400 pA. (c) A $-d^2I/dV^2$ spectrum obtained by differentiating the $dI/dV$ spectrum in panel (b).}
  \label{Fig3}
\end{figure}

\section{Results}

\begin{figure*}[t]
\includegraphics[width = 0.9\textwidth]{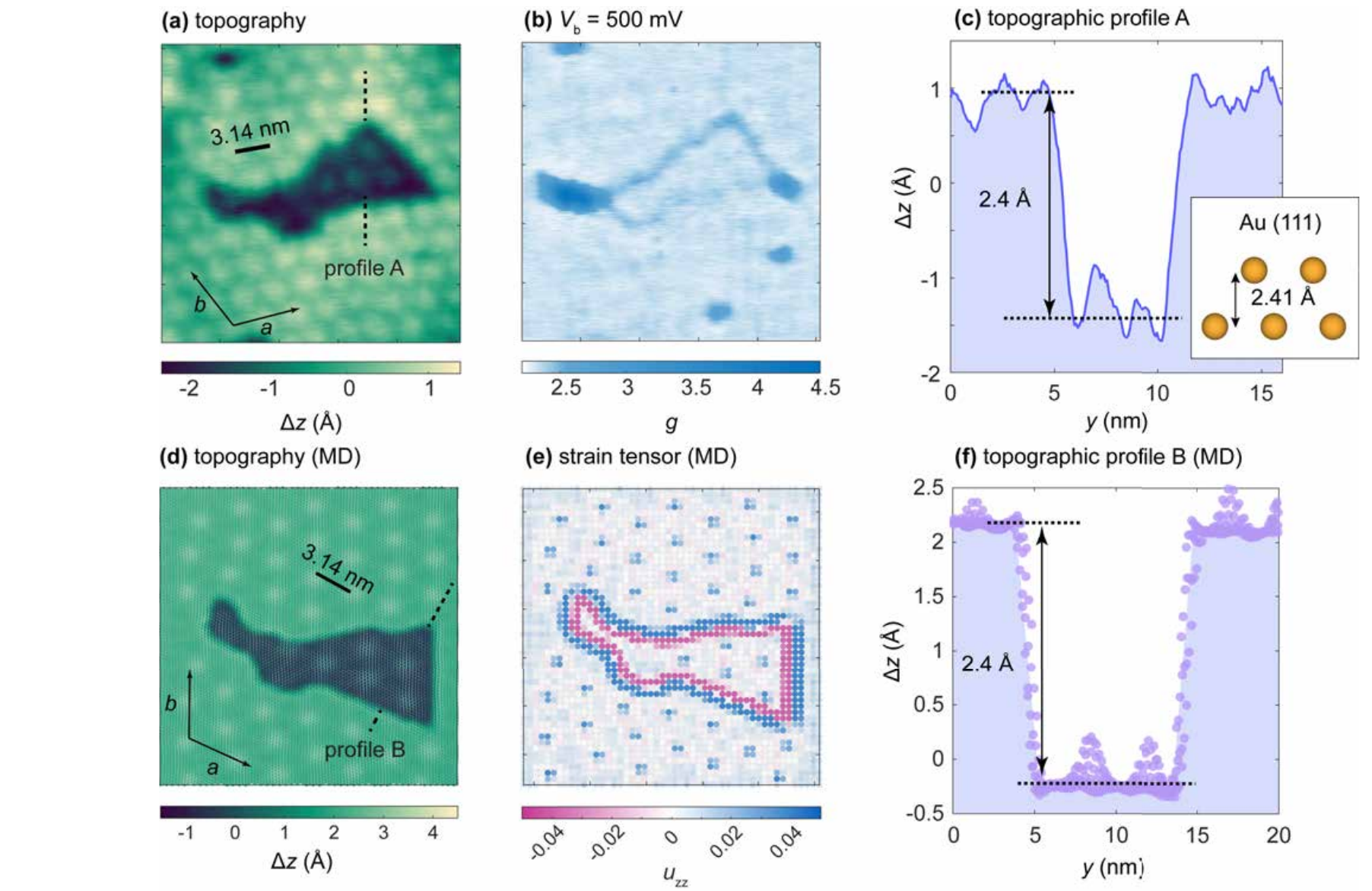}
  \caption {MoS$_2$ strained by nanoscale corrugations in Au(111) substrate: (a) An STM topographic map over a (30 nm $\times$ 30 nm) area encompassed by the white dotted line in Fig. 2(c). (b) A constant bias conductance map of the region of panel (a) with a set-point of $V_\mathrm{b} = 500$ mV and $I_\mathrm{t} = 500$ pA. (c) A topographic line profile obtained along the dashed black line indicated in panel 4(a). The inset indicates the cross section of the Au(111) lattice structure, where the distance between consecutive Au(111) planes is 2.41 \AA. (d) A MD simulated topographic map of the top S atom of a ML-MoS$_2$ interfaced on top of an Au(111) substrate with an indentation. (e) Strain map of the MD simulated monolayer MoS$_2$ on Au(111) exhibiting the $u_{zz}$ strain component. (f) A topographic line profile of the top S layer of MoS$_2$ across the indentation, showing good agreement with the topographic profile A measured by STM as depicted in panel (c).}
  \label{Fig4}
\end{figure*}

\subsection{Electronic properties of flat-area MoS$_2$-Au(111) heterostructure}

The procedures for the synthesis of ML-MoS$_2$ single crystals and the development of the MoS$_2$-Au(111) heterostructure for STM/STS studies are detailed in Methods. Owing to the nature of the Au evaporation process, the ML-MoS$_2$ single crystalline flakes are not laid over a perfectly flat Au(111) surface, but rather on a rugged Au surface with roughness of $\Delta z \approx 20$ nm over distance of several micrometers (Fig. \ref{Fig2}(a,b)). Consequently, like how a plastic wrap molds over an irregular surface, ML-MoS$_2$ wraps over the corrugations of the Au(111) grains and their terraces, thus forming wrinkles as well as Moir\'e patterns within a single flake, as shown in Fig. 2(c). Further examining a nanoscale area reveals a Moir\'e superlattice with a periodicity of 3.15 nm (Fig. \ref{Fig2}(e)). The observed periodicity in the Moir\'e pattern is highly consistent to values of those reported on MoS$_2$ grown directly on or transferred on Au(111) substrates, ranging from 3.15 to 3.3 nm \cite{Krane2016, Grønborg2015, C8NA00126J, Wu2020}, thus adding confidence to our notion that the Au substrate is mainly comprised of the (111)-lattice plane. Slight deviation of the Moir\'e lattice periodicity may signify a minor in-plane twist between the $(a, b)$-lattice vectors of ML-MoS$_2$ and those of the Au(111).

The theoretical electronic bandstructure of ML-MoS$_2$ and the representative differential conductance ($dI/dV$) and tunneling current ($I$) spectra taken by STM on a flat region of ML-MoS$_2$ on Au(111) across a range of bias voltages ($V_\mathrm{b}$) are shown in Figures \ref{Fig3}(a)-(c). The low conductance region in Figure \ref{Fig3}(b) representing the electronic bandgap ($E_\mathrm{g}$) is approximately 1.8 eV, consistent to that reported in Refs. \cite{Krane2016, Tumino2020, Park_2018} and the optically determined bandgap of 1.824 eV measured through the PL spectroscopy (Fig. \ref{FigS1}) for ML-MoS$_2$, and the monolayer nature of the investigated MoS$_2$ is further confirmed by the Raman spectroscopic study, as shown in Fig. \ref{FigS2}. The Fermi level ($E_\mathrm{F}$), which corresponds to $V_\mathrm{b} = 0$ in the tunneling spectra, is only $\approx$ 400 meV below to the conduction band minimum (at the K-valley), indicating that the ML-MoS$_2$ is an electron-doped semiconductor ($n$-type). Notably, the electronic local density of states (LDOS) exhibits a kink at 800 meV above $E_\mathrm{F}$, which corresponds to the bottom energy of the Q-valley conduction band (Fig. \ref{Fig3}(b,c)), where the Q-valley is located midway along the K-$\Gamma$ path in the first Brillouin zone (FBZ) (Fig. \ref{Fig3}(a)). Although excitations by $E_\mathrm{photon} = 2.41$ eV ($\lambda = 515$ nm) photons directly pump photoexcited QPs exclusively to the K-valley, scattering processes such as those through electron-phonon interactions will result in population of photoexcited QPs in the Q-valley as well.

\subsection{Strain-induced changes in the electronic structure}

As depicted in Fig. \ref{Fig2}(c), the ML-MoS$_2$ layer on Au(111) experiences strain not only from the periodic topographic modulations associated with the Moir\'e superlattices but also from the nanoscale corrugations in the Au(111) substrate. To delve into the impact of these corrugations on ML-MoS$_2$, we focus on a specific topographic indentation resembling a pit-like indentation feature within a (30 nm $\times$ 30 nm) area marked by white dashed lines in Fig. \ref{Fig2}(c), which is enlarged in Fig. 4(a). This analysis aims to comprehensively understand how these corrugations strain the ML-MoS$_2$. The presence of a Moir\'e superlattice inside the indentation, exhibiting the same periodicity as that in the flat regions of MoS$_2$-Au(111) heterostructure, suggests that this indented feature is not a result of a missing MoS$_2$ layer. Instead, it arises from a depression in the Au(111) substrate below, an observation also reported in Ref. \cite{Krane2016}. Additionally, Fig. \ref{Fig4}(c) reveals a height difference of 2.4 \AA$\hspace{1 pt}$ across the step of the indentation, which corresponds to the height of a single layer of Au(111).

With a clear understanding of the geometry of the indentation, we turn to Molecular Dynamics (MD) simulations to investigate the strain imposed on the ML-MoS$_2$ when wrapped over this indentation. Specifically, a multilayer Au(111) atomic slab structure, modeled based on the STM topographic image in Fig. \ref{Fig4}(a), was constructed. Through MD simulations, a MoS$_2$ monolayer was placed on top, resulting in a heterostructure as depicted in Fig. \ref{Fig4}(d). Notably, the Moir\'e periodicity of approximately 3.14 nm obtained through the MD simulations aligns well with our experimental observation. Moreover, the height difference of the top sulfur layer in MoS$_2$ across the pit from the MD simulations is 2.4 \AA, as shown in Fig. \ref{Fig4}(f), which is consistent with the STM findings (Fig. \ref{Fig4}(c)). This consistency thus supports the conclusion that the observed indentation via STM is a consequence of the corrugations in the Au layer underneath.

From the relaxed MoS$_2$-Au(111) heterostructure, the strain tensor is derived, which reveals that while there is not a significant difference in the in-plane strain tensor components (i.e., $u_{xx}$, $u_{yy}$ and $u_{xy}$) between the flat region and the indented region (Fig. \ref{FigS3}), the out-of-plane tensor component $u_{zz}$ exhibits pronounced values at the edge of the indentation as shown in Fig. \ref{Fig4}(e). The notable increase in $u_{zz}$ along the indentation edge indicates that, instead of loosely draping over the step, ML-MoS$_2$ undergoes a sharp lateral bend, tightly wrapping over the step. This behavior is also evident from the sharp drop in the STM topographic line profile at the edge (Fig. \ref{Fig4}(c)). Considering the impact of strain on the electronic structure, discernible alterations in the LDOS at the edge are expected. In line with this expectation, the STS map acquired at a constant bias of $V_\mathrm{b} = 500$ mV from the identical region reveals a larger LDOS at the edge of the indentation. This effect is even more apparent at the corner-like features of the indentation, as depicted in Fig. \ref{Fig4}(b) and Fig. \ref{FigS4}, conceivably due to a larger strain gradient over a nanometer area (see Fig. \ref{FigS5} for further discussion).

Expanding beyond the specific indentation highlighted, an examination of the topographic map encompassing a broader region, as illustrated in Fig. \ref{FigS6}, reveals that any indentations or ledges present are consistent with precisely either one or two multiples of the layer thickness within the Au(111) lattice. In general, we note that the strain is more pronounced at the corner-like features compared to the straight step-edges. As the corner-like feature becomes sharper, the $u_{zz}$ strain component intensifies and the LDOS is enhanced with sharper corner-like features among the conduction bands, as exemplified in Fig. \ref{FigS7} for the $u_{zz}$-strain component distribution maps and the DOS comparisons of corner-like features with angles of 90$^\circ$, 60$^\circ$, and 30$^\circ$. 

We emphasize that the distinctive topographic features shown in Figs. \ref{Fig2} and \ref{Fig4} do not stem from multilayer MoS$_2$, as thickness exceeding a single layer would have obscured the Moir\'e superlattice, leading to substantial alterations in the LDOS, as illustrated in Fig. \ref{FigS8}. Moreover, the monolayer nature of the investigated MoS$_2$ is further evidenced from PL and Raman spectroscopy, as shown in Figs. \ref{FigS1} and \ref{FigS2}. The observed topographic features therefore emanate from the corrugations in the substrate underneath. As the landscape of available energy states heavily impacts the behavior of photoexcited QPs, the effect of the strain-modified electronics states of ML-MoS$_2$ on its photoexcited stead-state LDOS is subsequently probed through STM while illuminating the sample with light.

\subsection{Photo-induced electronic bandstructure renormalization}

\begin{figure}[t]
\includegraphics[width = 0.5\textwidth]{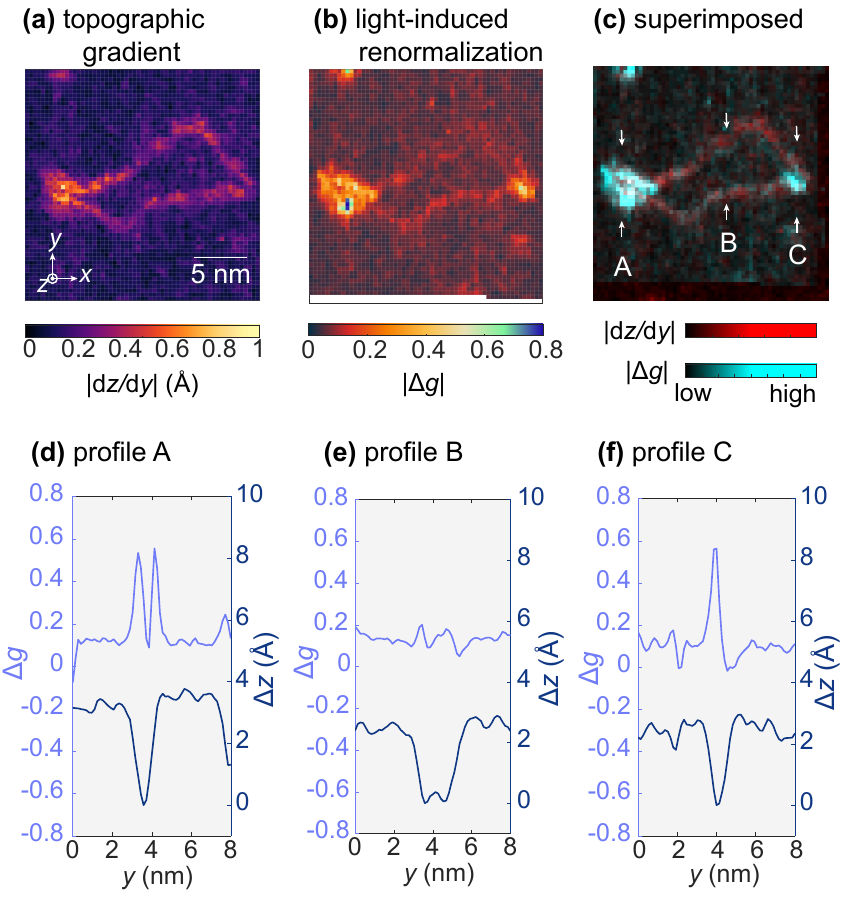}
  \caption {Enhanced light-induced renormalization at nanoscale strained region of ML-MoS$_2$-Au(111): (a) A topographic gradient map, ($dz/dy$), of the topographic map shown in Fig. \ref{Fig4} (a). (b) A $\Delta g(\vec{r},V_\mathrm{b})$ map representing the changes in the LDOS upon illumination of light measured at $V_\mathrm{b} = 600$ mV while maintaining a constant tunneling resistance of 1 $G\Omega$. (c) A superimposed image of panel (a) and (b), in which the arrows indicate lines where profiles cuts were taken. (d-f) Profile cuts (A, B, and C) from the $\Delta g(\vec{r},V_\mathrm{b})$ map in (b) and the topographic map in (a) taken along sections indicated by arrows in panel (c).}
  \label{Fig5}
\end{figure}

\begin{figure}[t]
\includegraphics[width = 0.45\textwidth]{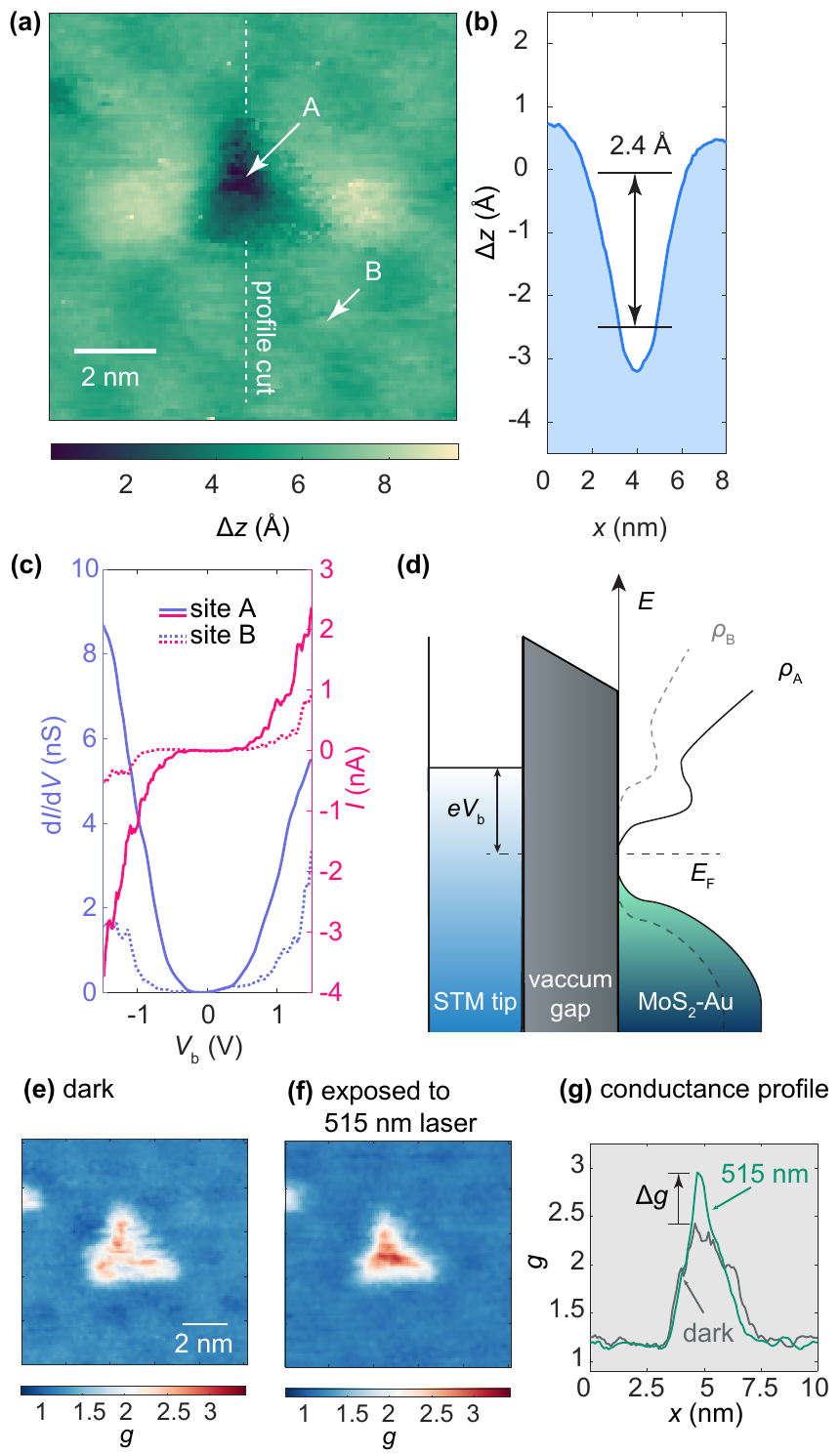}
  \caption {Reduced bandgap $E_\mathrm{g}$ and enhanced light-induced LDOS renormalization at nanoscale strained regions of monolayer MoS$_2$ on Au(111): (a) STM topographic image obtained with $V_\mathrm{b} = 800$ mV and $I_\mathrm{t}$ = 800 pA, showing a nanoscale triangular indentation in Au(111) in which MoS$_2$ is wrapped over. (b) The topographic line profile obtained from the dashed white line in panel (a). (c) $dI/dV$ and $I$ versus $V_\mathrm{b}$ spectra obtained from site A and site B, as indicated in panel (a). (d) A schematic diagram of MoS$_2$-Au(111) heterostructure junction depicting the strain-induced change in LDOS at site A and site B. Constant bias conductance map at $V_\mathrm{b} = 400$ mV and $I_\mathrm{t}$ = 400 pA (e) in the dark, and (f) under exposure to laser of a wavelength $\lambda = 515$ nm. (g) A linecut of the conductance profile taken from the constant bias conductance map of panels (e) and (f), depicting the change in LDOS upon illumination of light.}
  \label{Fig6}
\end{figure}

To systematically investigate the spatial variations in the light-induced electronic bandstructure renormalization of ML-MoS$_2$ on Au(111), measurements of constant bias conductance maps were carried out, where the ML-MoS$_2$ on Au(111) heterostructure was continuously illuminated with a laser power of 7 mW/cm$^2$ at varying STM bias voltages ($V_\mathrm{b}$). We note that the laser power used here is on the order of 10$^{-6}$ smaller than the laser power regime of kW/cm$^2$ necessary for sufficient heating to both broaden and red-shift the bandgap of ML-MoS$_2$ by tens of meVs \cite{Yu2019}, thus assuring the absence of any discernible light-induced heating effects in our investigation. Considering the substantial thermal conductivity of the 80 nm thick layer of Au that the ML-MoS$_2$ is in contact with, 317 Wm$^{-1}$K$^{-1}$ \cite{Lin2020}, even when assuming full conversion of all laser power to heat flux, the temperature at the surface of the ML-MoS$_2$ would have only experienced a negligible increase of $\approx 1.8 \times 10^{-8}$ K in temperature compared to the region not illuminated by the laser, further minimizing the relevance of light-induced heating effects in our experiments. Furthermore, when illuminating the tunneling junction with a laser, the formation of a metal/dielectric/metal (MDM) heterostructure by the junction consisting of STM tip/tunneling gap plus ML-MoS$_2$/Au substrate resulted in evanescently confined electromagnetic excitations at the interface and created surface plasmon polaritons (SPPs), as numerically verified through the Finite-Difference Time-Domain (FDTD) simulations using COMSOL Multiphysics (Fig. \ref{FigS9}) and experimentally demonstrated in Ref. \cite{Martín-Jiménez2020}. The confinement of SPPs to sub-wavelength volumes resulted in a $10^2 - 10^3$-fold local electric field enhancement (Fig. \ref{FigS9}), which intensified the light-matter interaction despite a relatively weaker laser power used in our experiment when compared with other optically detected light-induced band renormalization effects \cite{Ugeda2014, Cunningham2017, Pogna2016, Sie2017, Chernikov2015, PhysRevLett.122.246803, PhysRevB.106.L081117}.

In this experimental setup, the measured current across the STM junction under the illumination of light involves not only quasiparticle (QP) tunneling induced by the STM bias voltage modulation but also photoexcited carriers \cite{PhysRevLett.66.1717}, thus resulting in modifications to the LDOS of the illuminated sample. Under steady light illumination, the LDOS  $\rho_\mathrm{s,light}(\vec{r},V_\mathrm{b})$ of ML-MoS$_2$ due to photo-induced steady-state renormalization relative to the LDOS  $\rho_\mathrm{s,dark}(\vec{r},V_\mathrm{b})$  of ML-MoS$_2$ in the absence of light may be obtained from the changes in the normalized tunneling conductance by considering 
\begin{align}
\Delta g(\vec{r,V_\mathrm{b}}) &\equiv g_\mathrm{light}(\vec{r},V_\mathrm{b}) - g_\mathrm{dark}(\vec{r},V_\mathrm{b}) \notag \\
& \propto \rho_\mathrm{s,light}(\vec{r},V_\mathrm{b}) - \rho_\mathrm{s,dark}(\vec{r},V_\mathrm{b}),
\end{align}
where the normalized conductance at a given position $\vec{r}$ under a biased voltage $V_\mathrm{b}$ is empirically obtained by 
\begin{align}
\Delta g_\alpha(\vec{r,V_\mathrm{b}}) &\equiv \frac{V_\mathrm{b}}{I_\alpha (\vec{r,V_\mathrm{b}}) } \cdot \frac{dI_\alpha (\vec{r},V_\mathrm{b})}{dV}\Bigg|_{V = V_\mathrm{b}},
\end{align}
where the subscript $\alpha$ denotes either the light or dark condition. 

\begin{figure*}[t]
\includegraphics[width = 0.9\textwidth]{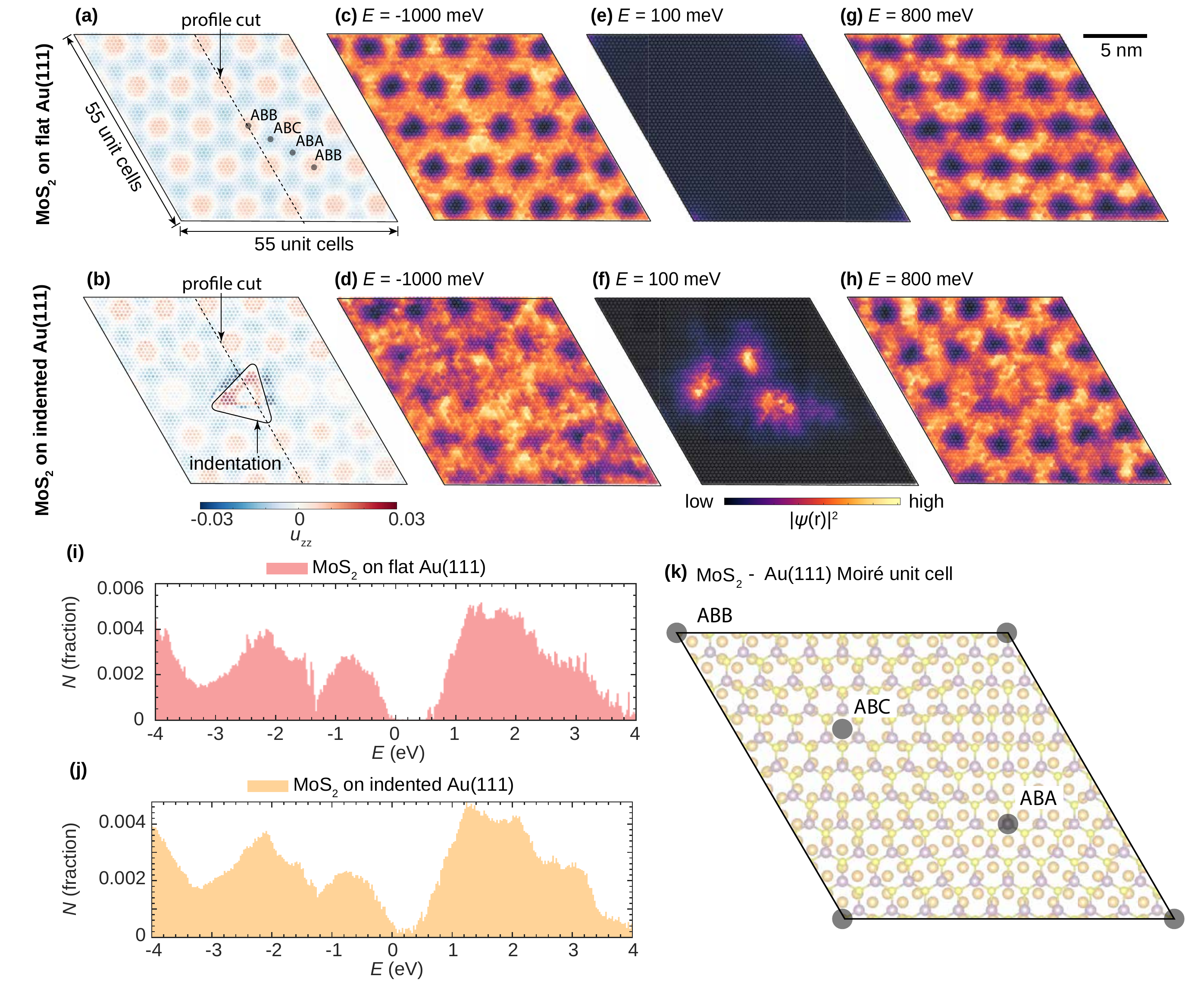}
  \caption {Calculated strain-induced modification on the electronic structure: The $u_{zz}$ strain tensor of a (55 $\times$ 55) finite ML-MoS$_2$ lattice laid over (a) a flat Au(111) surface and (b) an indented Au(111) surface. The calculated electron density $\rho(E, \vec{r}_i)$ maps at various energies for MoS$_2$ on flat Au(111) and indented Au(111) are shown in (c,d) for an energy below the valence band ($E = -1000$ meV), in (e,f) for a mid-gap ($E = 100$ meV), and in (g,h) for an energy above the conduction band ($E = 800$ meV). Each panel shares the same color codes shown in the color bar but have differing color scales. The global DOS versus energy ($E$) spectra of the entire (55 $\times$ 55) lattice shown in (i) for ML-MoS$_2$ on flat Au(111) and in (j) ML-MoS$_2$ on indented Au(111), revealing an overall reduced energy gap for the strained ML-MoS$_2$ due to the indentation. (k) A diagram of a Moir\'e unit cell of MoS$_2$ -Au(111) heterostructure consisting of 10 $\times$ 10 MoS$_2$ unit cells and 11 $\times$ 11 Au(111) unit cells, depicting the ABB, ABC and ABA stacking domains. Here the yellow, purple, and orange circles denote the S, Mo, and Au atoms, respectively. }
  \label{Fig7}
\end{figure*}

Figure \ref{Fig5} illustrates our quantitative analysis of the light-induced LDOS renormalization effects on ML-MoS$_2$ as a function the biased voltage $V_\mathrm{b}$ through mapping the changes in the normalized tunneling conductance $\Delta g(\vec{r},V_\mathrm{b})$ at each pixel $\vec{r}$ upon illumination of light. Comparing the topographic gradient ($dz/dy$) map in Fig. \ref{Fig5}(a) with the light-induced LDOS renormalization map in Fig. \ref{Fig5}(b), we note that substantial changes in $\Delta g(\vec{r},V_\mathrm{b})$ upon exposure of light primarily appear in regions with pronounced strain, as depicted in Fig. \ref{Fig5}(c). It is evident that regions at the edge of the indentation exhibit a pronounced light-induced renormalization effect compared to the flat region. This finding emphasizes the localized effect of light illumination on renormalizing the electronic structure of ML-MoS$_2$ primarily in nanoscale regions with proximity to the indentation boundary that also exhibits substantial strain-induced bandstructure modifications.

\subsection{Modeling nanoscale strain-induced modulations in the energy landscape }

\begin{figure*}[t]
\includegraphics[width = 0.95\textwidth]{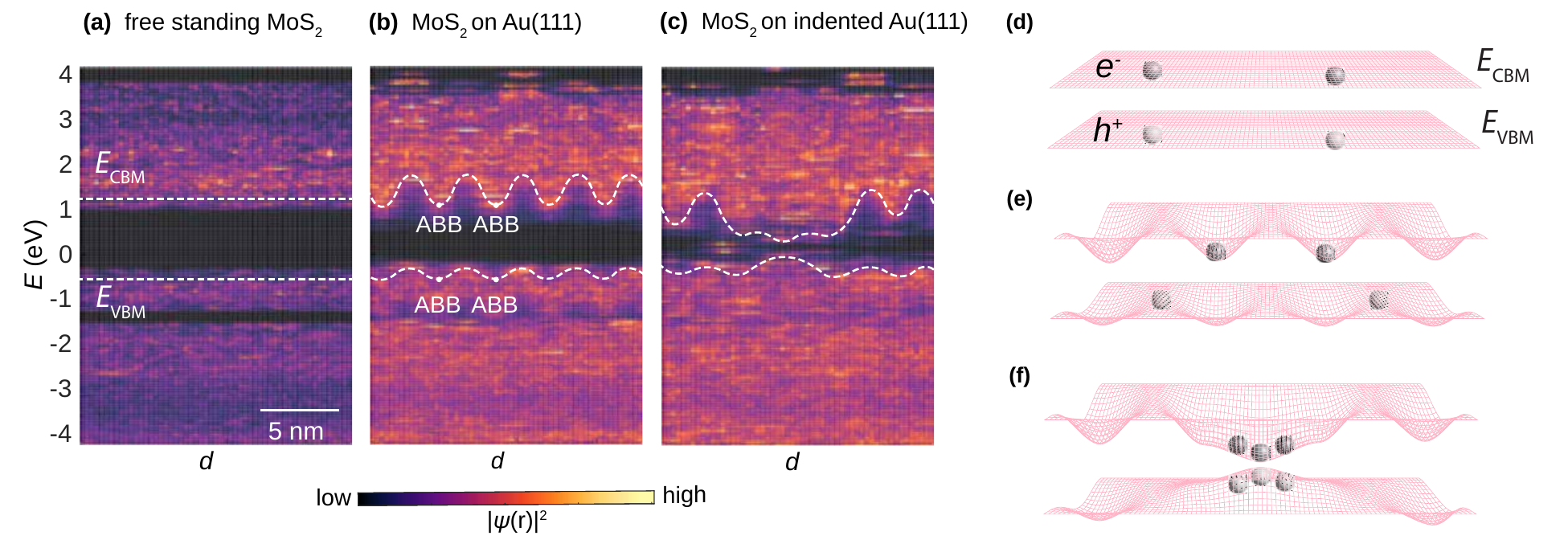}
  \caption {Effects of strain on electron density distribution: The calculated energy- dependence of $\rho(E,\vec{r}_i)$ taken along a profile cut in the middle of the (55 $\times$ 55)-finite lattice for (a) a free-standing ML-MoS$_2$, (b) a ML-MoS$_2$ on flat Au(111), and (c) a ML-MoS$_2$ on indented Au(111). Schematic diagrams indicating the spatial energy landscape of the conduction band and the valence band are shown in (d) for free-standing MoS$_2$, in (e) for MoS$_2$ on flat Au(111), and in (f) for MoS$_2$ on indented Au(111). The circles represent the photoexcited hole ($h^+$) and electron ($e^-$) carriers.}
  \label{Fig8}
\end{figure*}

\begin{figure*}[t]
\includegraphics[width = 0.9\textwidth]{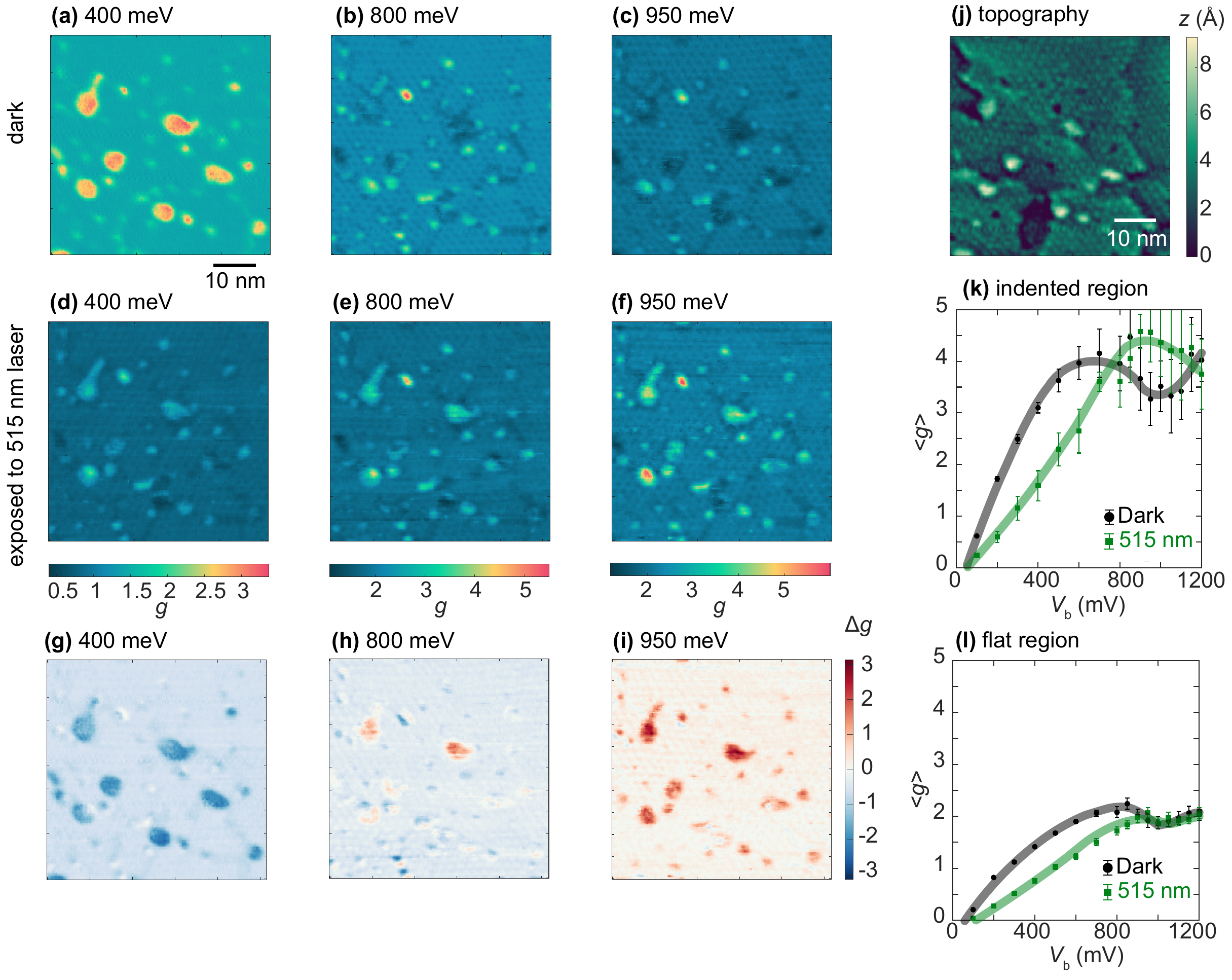}
  \caption {Bias-dependence of light-induced renormalization of LDOS: Constant bias normalized conductance map, $g(\vec{r},V_\mathrm{b})$, obtained at various bias voltages (a-c) without the illumination of light, and (d-f) under the illumination of 515 nm light, measured while maintaining a tunneling resistance of 1 G$\Omega$. A line noise artifact removal was performed to enhance contrast. (g-i) The difference in the normalized conductance map, $g(\vec{r},V_\mathrm{b})$, taken with and without the illumination of light at different biases ($V_\mathrm{b}$). (j) Topographic map of the areas corresponding to the conductance maps measured with a set point of $V_\mathrm{b}$ = 800 mV and $I_\mathrm{t}$ = 800 pA. (k) The bias dependence of the average g at the indented region, and (l) at the flat region. The error bars indicate the first standard deviation, and a best fit line is drawn to depict the trend.}
  \label{Fig9}
\end{figure*}
To further delve into the observed enhancement of light-induced bandstructure renormalization in nanoscale strained regions, we specially focus on a single nanoscale indentation in Au, as illustrated in Fig. \ref{Fig6} (a), for its simple triangular geometry. Studying such elementary geometry facilitates straightforward modeling and simulation for comparison with the experimental results. Given that the topographic indentation is $\approx 2.4$ \AA$\hspace{1 pt}$ (Fig. \ref{Fig6}(b)), the indentation here is owing to a single layer indentation of Au(111). The ML-MoS$_2$ strained by this triangular indentation displays a $dI/dV$ spectrum (site A) with a larger LDOS and a narrower zero-conductance region close to $E_\mathrm{F}$ in comparison to the point spectrum in the flat region (site B) shown in Fig. \ref{Fig6} (c), suggesting that there is a reduction of the bandgap at the nanoscale strained region (Fig. \ref{Fig6} (d)). Moreover, similar to the observation in Fig. \ref{Fig5}, while negligible differences in  $\Delta g(\vec{r},V_\mathrm{b})$  are observed in the flat regions outside of the indentation for laser light of a wavelength $\lambda = 515$ nm, a notable contrast in $\Delta g(\vec{r},V_\mathrm{b})$ under the same photoexcitation emerges within the highly strained region of the indentation (Fig. \ref{Fig6} (e,f)).

The electronic structure of ML-MoS$_2$ over this triangular indentation observed in Fig. \ref{Fig6}(a) is modelled by constructing a (55 × 55)-lattice ML-MoS$_2$ laid over a triangular Au indentation geometry of the same scale, which is compared with a regular (55 × 55)-lattice without indentation. While a regular periodic Moir\'e pattern is evident for the case of MoS$_2$ placed over flat Au(111) surface without the presence of indentation (Fig. \ref{Fig7}(a) and Fig. \ref{FigS10}), the presence of a triangular indentation results in a strain tensor field exhibited in Fig. \ref{Fig7}(b) and Fig. \ref{FigS11}, where the translational invariance of the Moir\'e periodicity is clearly destroyed over multiple periods. 
 
With the strain levels induced by the indentation spanning $\approx \pm 3\%$ (Fig. \ref{Fig7}(b)), significant alterations are expected to occur in the local electronic states, which we confirm in Fig. \ref{FigS12} that displays the comparison of the global DOS of the entire ($55 \times 55$)-lattice for free-standing ML-MoS$_2$, ML-MoS$_2$ on flat Au(111), and ML-MoS$_2$ on Au(111) with the triangular indentation. This finding is also consistent with calculations of the modified electronic bandstructure under uniform strain, as systematically illustrated in Fig. \ref{FigS13}, although we note that due to the spatially non-uniform nature of the strain tensor field resulting from the indentation, conventional considerations in terms of bandstructure in momentum space are insufficient.

For realistic comparison with experimental observation of spatially varying strain, our first principle calculations employ a real-space superlattice Hamiltonian, where the strain tensor field is applied by modifying the hopping integral through changes in bond length and bond angle that are calculated using the computed Green-Lagrange strain tensors, as detailed in the notes leading to Fig. \ref{FigS14} in the Supporting Information. This approach enables the observation of strain-induced spatial modulations in the electronic states, as detailed in Methods. The alterations in the energy bands due to non-uniform strain in the real-space finite lattice Hamiltonian are investigated by diagonalizing the Hamiltonian to yield its eigenvalues ($E_n$) and eigenstates ($\psi_n (\vec{r})$), as described in Methods under the subsection \textit{First principle calculations}.  In this computation, the impact of the Moir\'e superlattice on the electronic structure of ML-MoS$_2$ is considered by introducing a periodic strain potential, whereas other Moir\'e-induced effects such as interlayer vdW tunneling with the underlying Au atoms are neglected. This approximation is justifiable because the vdW interaction energy between MoS$_2$ and the substrate Au atoms only yields local chemical potential modulations on the order of 10 meV \cite{PhysRevB.93.165422, Geng2020, Silva_2022}, while the influence of strain-induced topographic displacement alters the electronic overlap integrals on the order of 1 eV (Fig. \ref{FigS14}), thus being more predominant.

Figure \ref{Fig7}(i) displays a histogram of En, which represents the energy spectrum of the global density of states (DOS) for ML-MoS$_2$ on flat Au(111). In contrast to the DOS of a free-standing ML-MoS$_2$ (Fig. \ref{FigS12} (a,b)), once interfaced with Au(111), Moir\'e superlattice-induced strain potential is prompted, resulting in an increase in valence band maxima ($E_\mathrm{VBM}$) and a decrease in conduction band minima ($E_\mathrm{CBM}$), simultaneously blurring features in the global DOS. Such features are further blurred upon implementation of the triangular indentation as shown in Fig. \ref{Fig7}(j), and the zero-DOS region near $E_\mathrm{F}$ fully diminishes. This reduction in the zero-DOS region reflects the modulation of eigenstates not only by the Moir\'e potential but also by the presence of the indentation.

For further spatial analysis of the electron distribution at a given energy $E$, we examine the probability density  $\rho(E,\vec{r}_i) = |\Psi_i(E,\vec{r}_i)|^2$. While evaluation of DOS provides information about the number of available states of the Hamiltonian at a specific energy window, the electron probability density provides the spatial distribution of electrons within those states. Here, $\Psi_i$ represents a superposition of the basis functions $\psi_n(E,\vec{r}_i)$ that fall within an energy window ($2\delta$) centered at a specific energy $E$ at each orbital position, $\vec{r}_i$, which is given by

\begin{align}
\ket{\Psi_i(E,\vec{r}_i)} &= \sum _{E - E_n < \delta} \ket{\psi_n(E,\vec{r}_i)}.
\end{align}

Using an energy window of $2\delta = 50$ meV for our analysis, Figs. \ref{Fig7}(c,e,g) depict the probability density distribution $\rho(E,\vec{r}_i)$ at various energies for ML-MoS$_2$ on a flat Au(111) substrate. Following the convention of Ref. \cite{KRANE2018136}, a single Moir\'e unit cell consists of three stacking domains: where the S atom sits right on top of the Au atom (ABB stacking) or on a hollow-site (ABC stacking), and where Mo atom sits right above the Au atom (ABA stacking), as shown in Fig. \ref{Fig7}(k). A distinct difference in $\rho(E,\vec{r}_i)$ across the stacking domains is evident in energies above $E_\mathrm{CBM}$ and below $E_\mathrm{VBM}$. 

Upon introducing an indentation in the underlying Au(111) layer, the Moir\'e patterns in the electron density exhibit reduced periodic order (Fig. \ref{Fig7}(d,f,h)). While the Moir\'e pattern-induced periodic electron density is still observable, such periodic patterns are notably absent at the location of the indentation. There is a higher electron density where the indentation is located, which is consistent to the experimental LDOS data through the $dI/dV$ spectral measurements (Fig. \ref{Fig6}(e)). The enhanced LDOS is present at both states above the conduction band minimum and below the valence band maximum. To evaluate the energy dependence of the electron density distribution, a profile cut was performed, which traversed the middle of the crystal, through the dashed line depicted in Fig. \ref{Fig7}(a,b).

In the case of a free-standing ML-MoS$_2$, the $E_\mathrm{CBM}$ and $E_\mathrm{VBM}$ are spatially flat (Fig. \ref{Fig8}(a)). Consequently, there are no energetically favorable energy minima where the photoexcited carriers or excitons can drift towards, leading to an expected equal distribution of photoexcited QPs throughout the crystal. However, by placing MoS$_2$ on top of flat Au(111), Moir\'e-induced modulations in the electronic states emerge (Fig. \ref{Fig8}(b)). The modulation amplitude of $\approx 0.7$ eV in the conduction bands is more pronounced than that in the valence bands ( $\approx 0.2$ eV), consistent with experimental STM observations from conductance maps measured at various biases (Fig. \ref{FigS15} and Fig. \ref{FigS16}). Specifically, the ABB stacking site attains a $E_\mathrm{CBM}$ lower in energy, and a $E_\mathrm{VBM}$  higher in energy relative to all the other stacking domains.

As photoexcited electrons in the conduction band settle into the valleys of the modulated energy landscape, and holes residing in the valence bands are drawn to the apex, a non-homogeneous spatial distribution of photoexcited carriers is expected. Given that the exciton binding energy in MoS$_2$-Au(111) heterostructures is approximately 90 meV \cite{Park_2018}, the substantial energy gradient provided by the Moir\'e-induced energy modulations is capable of disassociating the excitons into spatially separated electrons and holes \cite{Feng2012}, as illustrated in Fig. \ref{Fig8}(e).

With the indentation in the lattice, on the other hand, ECBM at the center of indentation is lowered, and $E_\mathrm{VBM}$ is raised, resulting in a reduction of the bandgap at the center of this defect to approximately $E_\mathrm{g} \approx 0.5$ eV (Fig. 8(c)), which opens a channel for photoexcited QPs, including excitons and electron-hole plasmas, to migrate towards this indented region (Fig. \ref{Fig8}(f)). As the energy gradient is spread over a length of a few nanometers, comparable to the mean-free path of electrons in MoS$_2$ \cite{Guo_2019, Smithe_2017}, charge carriers are capable of drifting into the strained region \cite{Moon2020, Shao2022}. Figure \ref{Fig6}(c), displaying the experimental $dI/dV$ spectra measured at the indented region and the flat region, indeed verifies the behavior of the calculated electronic states impacted by strain. While among the flat region, a bandgap of $E_\mathrm{g} \approx 1.8$ eV is revealed, the spectrum measured at the center of the indentation displays a significantly reduced bandgap of $E_\mathrm{g} \approx 0.7$ eV. Given the reduced energy gap, more states are expected at energies above $E_\mathrm{CBM}$ and below $E_\mathrm{VBM}$, as observed in the enhanced LDOS at the strained region in the constant bias conductance maps (Fig. \ref{Fig6}(e,f)). Thus, in these strained regions, photoexcited electron and hole pairs precipitate, creating a pocket to host higher densities of QP populations.

Considering that the indentation geometry can modify the local electronic states, the extent of light-induced renormalization may vary for different indented configurations. Moreover, the magnitude of changes in the local electronic states tends to increase with higher densities of the light-induced QPs, as indicated by previous studies.13 Hence, we may gain insights into the averaged collective effect arising from a multitude of topographic corrugations by examining the strain and light-induced renormalization effects on the electronic structure of ML-MoS$_2$ over a sufficiently large spatial area. Although topographic features due to the nanoscale corrugations in Au are a collection of indentations and protrusions, both are structural distortions with their strain effects on MoS$_2$ solely determined by the gradients of the local displacement fields. This analysis thereby provides information about the overall light-induced renormalization effect across various indented configurations and provides a comprehensive perspective on the cumulative effects of excitonic interactions with strain.
Figure \ref{Fig9}(a-f) displays the constant bias normalized conductance maps at various biases across a (50 nm $\times$ 50 nm) area, revealing inhomogeneous LDOS due to numerous nanoscale topographic features (Fig. \ref{Fig9}(j)). Across all bias voltages, the light-induced electronic bandstructure renormalization effect is accentuated in regions with topography-induced strain (Fig. \ref{Fig9}(g-i)). At lower biases, such as the constant bias conductance map shown in Fig. \ref{Fig9}(g) for $V_\mathrm{b} = 400$ mV, light illumination leads to a reduction in LDOS in the strained regions. Interestingly, at a higher bias of $V_\mathrm{b} = 800$ mV, there are strained regions where the LDOS either increases or decreases due to renormalization. Further increasing the bias to $V_\mathrm{b} = 950$ mV, the LDOS of strain regions experiences a renormalization to a higher value. Clearly, the renormalization effect on the LDOS is not the same throughout all bias voltages. To better understand the trend, we examine the $V_\mathrm{b}$-dependence in smaller bias increments, distinguishing between regions with enhanced LDOS due to strain and flat regions with smaller strain.

Figures \ref{Fig9}(k,l) illustrate the bias dependence between 200 mV and 1050 mV of the average conductance, $\braket{g(\vec{r},V_\mathrm{b})}$, for the region strained by the topographic features and the flat region, respectively. Here, the pixels of either the regions strained by the topographic features or the flat regions in the 2D map were classified by taking the norm of the numerically computed gradient vector ($\vec{\nabla} z (x,y)$) of a pixel and then labeling the pixel as residing in the strained region if the $|\vec{\nabla} z|$ value surpasses the mean $|\vec{\nabla} z|$ value of the entire map. As anticipated, the magnitude of the average LDOS is significantly larger in the strained region compared to that in the flat region. Additionally, in all cases, there is a non-monotonic bias dependence in the LDOS for both the indented and flat regions. In the flat region, a feature resembling a local maximum in the absence of light illumination is observed at $\approx 800$ mV, which may be attributed to the energy minimum associated with the Q-valley, as shown in the $dI/dV$ spectrum in Fig. \ref{Fig3}(b,c). This Q-valley feature in the strained region is, on average, shifted to a lower bias of $\approx 700$ mV, which is consistent with the overall downward shift of electronic states induced by strain, confirmed by theoretical calculations (Fig. \ref{Fig8}(c)) and by the empirical $dI/dV$ spectrum of Fig. \ref{Fig6}(c). Upon light illumination, the average LDOS distribution shifts to a higher energy, positioning the Q-valley feature at $\approx 900$ mV, which is more than 100 mV higher than the energy position of the Q-valley without light illumination. Interestingly, within the flat region, there is a small downward shift in the LDOS among lower biases (Fig. \ref{Fig9}(l)), and the overall LDOS alteration is much less than that of the strained regions. The reduction in the LDOS within the flat region in this bias voltage range upon light illumination implies that the lower-energy states may have been redistributed to energies outside of this bias window.

\subsection{Laser power dependence of the band renormalization}

\begin{figure*}[t]
\includegraphics[width = \textwidth]{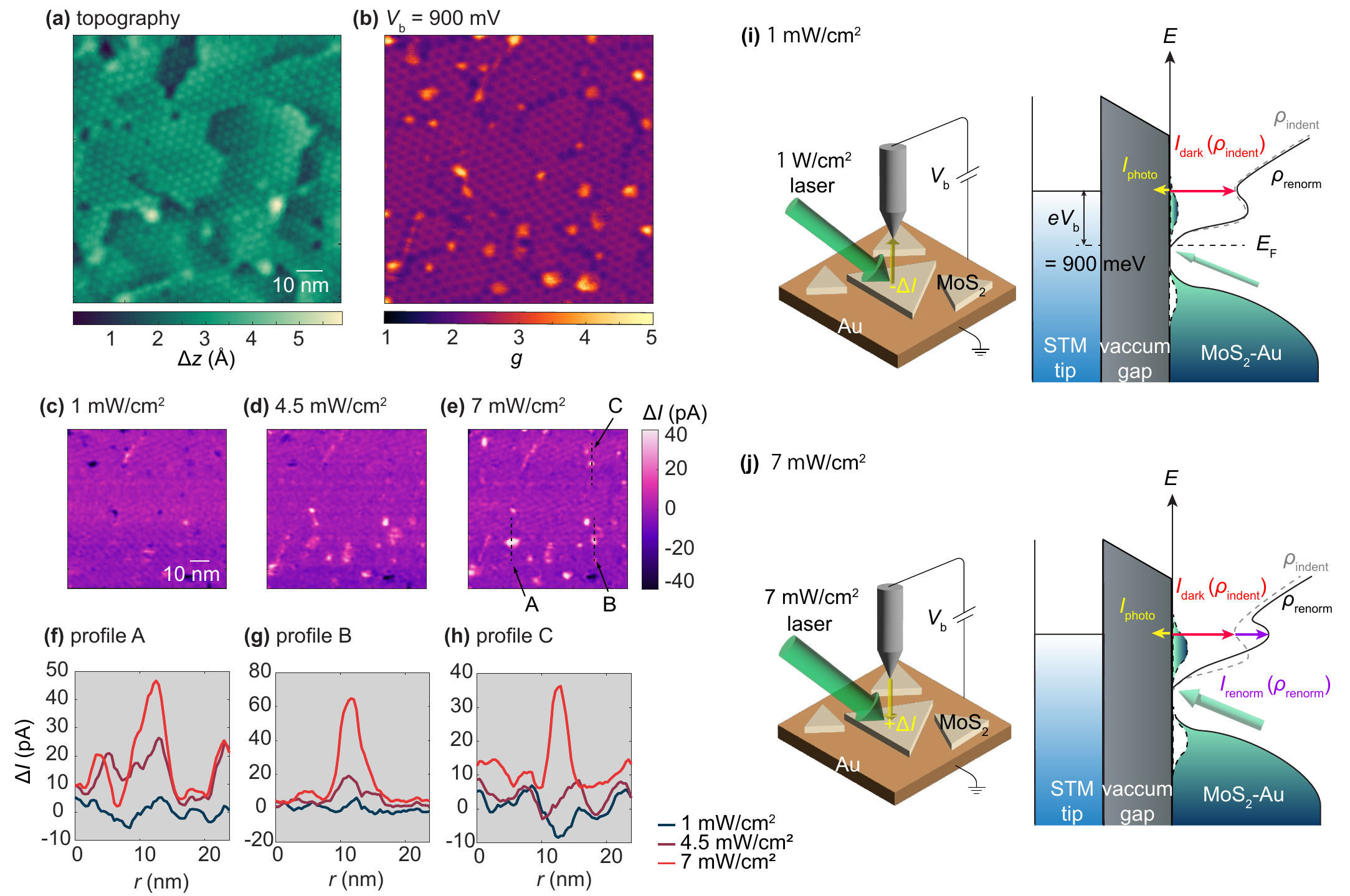}
  \caption {Laser power dependence of tunneling current: (a) A topographic map measured with a set-point of $I_\mathrm{t} = 900$ pA and $V_\mathrm{b} = 900$ mV. (b) Normalized conductance map measured at $V_\mathrm{b} = 900$ mV for the same area shown in panel (a). $\Delta I (\vec{r},V_\mathrm{b})$ maps under exposure of laser with (c) 1 mW/cm$^2$, (d) 4.5 mW/cm$^2$, and (e) 7 mW/cm$^2$. Profiles of $\Delta I$ taken as line cuts across indented sites A (f), B (g), and C (h) marked in (e). Schematic diagram indicating the direction of $\Delta I (\vec{r},V_\mathrm{b})$ upon the illumination of (i) low laser power and (j) high power laser along with schematic diagrams of the energy dependence of LDOS depicting the STM tip and MoS$_2$-Au(111) heterostructure junction in which the STM tip is biased to +900 meV for various experimental conditions.}
  \label{Fig10}
\end{figure*}

We have found that illumination of the sample with a 515 nm laser power of 7mW/cm$^2$ results in the LDOS associated with the Q-valley in the indented region to shift upwards to 900 meV, as shown in Fig. \ref{Fig9}(k). At this corresponding bias voltage, we further investigated the dependence of light-induced renormalization effect on the laser power. This study was performed by obtaining the current map measured by a lock-in amplifier (LIA) at each pixel with ($I_\mathrm{light}$) and without ($I_\mathrm{dark}$) exposure to light, and a convention for electrons tunneling from the STM tip to the sample as positive currents is utilized. By taking the difference in the measured tunneling current, 
\begin{align}
\Delta I(V_\mathrm{b}) = I_\mathrm{light} - I_\mathrm{dark},
\end{align}
information about both the effect of light on the LDOS and the occupation of the QP states was obtained. In the case of $\Delta I<0$, a current of electrons is directed from the sample to the tip, while a positive current  $\Delta I>0$  signifies a current of electrons from the tip to the sample (Fig. \ref{Fig10}(i,j)). In particular, negative tunneling current ($\Delta I<0$) is associated with the excess photocurrent arising from the photoexcited electrons in the conduction band of the sample tunneling to the tip (Fig \ref{Fig10}(i)) \cite{Fan1993, 10.1116/1.589571, 10.1063/1.1432113, 10.1063/1.41425, PhysRevB.53.8090, Li2024}. On the other hand, given that laser illumination increases the LDOS at $V_\mathrm{b} = 900$ mV, a current of electrons from the tip the sample is a manifestation of the increase in the LDOS ((Fig. \ref{Fig10}(j))) of the ML-MoS$_2$ due to light-induced renormalization effect. Figures \ref{Fig10}(c-e) display $\Delta I (\vec{r},V_\mathrm{b})$ distribution for various laser powers, measured at the same region in which its topographic map and a constant bias conductance map at$V_\mathrm{b} = 900$ mV are displayed in Fig. \ref{Fig10}(a,b). With exposure to a laser power of 1 mW/cm$^2$, the flat region of the sample shows a negligible degree of change, $\Delta I \approx 0$, except regions of indentation where patches of $\Delta I < 0$ are observed (Fig. \ref{Fig10}(c)), with the magnitude of the current reaching 10s of picoamperes. The occurrence of excess electron tunneling from the sample to the tip under forward voltages underscores the pronounced tendency of photoexcited carriers to concentrate in regions with strain within the MoS$_2$. Notably, given the substantial magnitude of the photoexcited QP drift length on the order of $10^2$ nm at room temperature \cite{Feng2012}, it provides a channel for excited carriers to effectively drift towards and aggregate at the nanoscale strained regions \cite{Moon2020, C7NR03537C}.

By increasing the laser power to 7 mW/cm$^2$, the indented regions display more patches with $\Delta I > 0$, which is as expected due to increased light-induced renormalization effects on the LDOS of ML-MoS$_2$ as a consequence of strong many-body interactions due to aggregation of QPs in nanoscale potential wells, particularly exciton polaritons with binding radii of several nanometers. The overall laser power dependence is visible by taking profiles at selected indented sites (Fig. \ref{Fig10}(f-h)). The non-monotonic power dependence of the LDOS at the strained region further discerns the observed light induced effect from a heating effect: Had such electronic renormalization been a simple heating effect, a larger laser power that induces more heating would have monotonically increased the global renormalization effect on the LDOS of the sample, which was contrary to the observed localized effect on the LDOS with increasing laser power only on nanoscale strained regions of the sample. Thus, our studies of the laser power dependence of the LDOS aligned well with the scenario of QP-induced renormalization effects in strongly strained regions.

\section{Discussions}

Given that strain in the indented region lowers the local conduction band and lifts the local valence band and that the finite lifetime of excitons allows for sufficient migration before electron-hole recombination, photoexcited excitons besides unbound electrons and holes may funnel toward the indented region. Noting that the average spacing between neighboring indentations are typically on the order of a few tens of nanometers, which is smaller than the exciton drift length of hundreds of nanometers \cite{Feng2012}, funneling of excitons into strained indentations is fully expected. Furthermore, the maximum laser power utilized in this study, at 7 mW/cm$^2$, is orders of magnitude smaller than the excitation power density required to transition into the Mott regime (kW/cm$^2$) where excitons fully dissociate into an electron-hole plasma \cite{PhysRevLett.122.246803, Yu2019}. Therefore, the renormalization effect observed in this work may be primarily attributed to excitons rather than an electron-hole plasma. Moreover, excitons are not mutually exclusive so that strong many-body interactions are expected due to their wavefunction overlaps in spatially highly confined regions, given the comparable nanoscale dimensions of the exciton Bohr radii \cite{PhysRevB.86.241201, PhysRevLett.113.076802, PhysRevLett.113.026803, Feng2012, PhysRevLett.111.216805} and strain-induced trapping potentials. We may further eliminate the possibility of either the optical Stark effect or the Bloch-Siegert effect contributing to the pronounced light-induced LDOS renormalization in the nanoscale strained regions of ML-MoS$_2$, as the laser power used here is exceedingly low as noted earlier, and there is no discernible hybridization of the conduction band visible in the $dI/dV$ spectra \cite{Sie2015, doi:10.1126/science.aal2241}.

An exciton-driven upward shift in the conduction band, similar to the findings in this investigation, was also observed in ML-MoS$_2$ via ARPES measurements reported in Ref.\cite{PhysRevB.106.L081117}. On the other hand, these results associated with ML-MoS$_2$ are in contrast to the light-induced renormalization effect on the Q-valley observed in ML-WS$_2$ on graphite \cite{Chen2022} and in ML-WS$_2$ on quartz \cite{Chernikov2015}, where the latter exhibited a downward shift of the conduction band upon photoexcitation. Moreover, while a theoretical calculation by Ref. \cite{PhysRevB.98.035434} of excitons-induced band renormalization predicts a general lowering of conduction bands, another calculation in Ref.\cite{PhysRevB.106.L081117} predicts an upward energy shift in the conduction band due to excitons. In any case, the nanoscale strain-induced modifications of the electronic structure observed in this study clearly cause a light-induced renormalization trend differing from those reported and/or predicted previously \cite{Chen2022, Chernikov2015, PhysRevB.98.035434}, the latter did not consider the light-induced renormalization effect with the presence of strain. We speculate that the pronounced light-induced shift in the Q-valley band is due to strain-induced lowering the energy of the conduction band at the Q-valley relative to the K-valley, consequently attaining an increased density of QPs in the Q-valley and leading to stronger renormalization effects. On the other hand, whether differences between the bandstructures of ML-MoS$_2$ and ML-WS$_2$ may contribute to the light-induced upward versus downward shift in the Q-valley conduction band remain an open issue for further investigation.

To gain further insight into the light-induced effects on the electronic structure of ML-MoS$_2$, we investigated the spectral dependence on the photon energy by illuminating the same ML-MoS$_2$ with a 635 nm ($E_\mathrm{photon} = 1.95$ eV) laser of equivalent power so that the photoexcited electrons could only populate the K-valley of flat free-standing ML-MoS$_2$, as depicted in Fig. \ref{FigS17}. Even for strained ML-MoS$_2$, the photoexcited carriers could only populate lower energy bands by illumination of 635 nm laser, thus limiting their diffusion range throughout the modulated energy landscape. As a result, we found that photoexcitation by 635 nm laser did not replicate the same notable light-induced renormalization effects with a 515 nm laser ($E_\mathrm{photon} = 2.41$ eV), as shown in Fig. \ref{FigS18}. The reduced photon-induced renormalization effect for illumination with light of $\lambda = 635$ nm may be attributed to the reduced photon energy, which resulted in lower-energy photoexcited QPs that were more susceptible to inelastic scattering, rendering them far less efficient in funneling across the modulated energy landscape to the nanoscale-strained regions. This finding further supports the notion that significant light-induced electronic bandstructure renormalization requires strong interactions of light-induced QPs, which may be achieved by enhancing QP funneling into local potential energy minima through nanoscale engineering of the ML-TMD material as well as by applying higher frequency and higher intensity of photoexcitation.      

\section{Conclusion}
In conclusion, we have demonstrated strongly enhanced light-induced electronic structure renormalization effects in nanoscale strained regions of ML-MoS$_2$ on corrugated Au(111) substrate through STM measurements under the illumination of $ \lambda= 515$ nm laser. This renormalization effect resulted in an upward energy shift of the Q-valley in the conduction band of ML-MoS$_2$ by more than 100 meV, which is significantly larger than the light-induced effect observed in flat non-strained ML-MoS$_2$ on graphite13 and in ML-WS$_2$ on graphite \cite{Chen2022}. Such strong light induced renormalization effect observed in this investigation is attributed to efficient aggregation of photoexcited QPs, particularly bosonic exciton polaritons, to nanoscale strained regions of the sample, as evidenced from our detailed STM measurements. While there have been previous investigations of exciton funneling into strained TMDs with micron-scale structures, considering the nanometer-scale exciton binding length, it is apparent that the nanoscale strain-induced confinement potentials in this work result in much stronger photoexcited QP interactions and therefore intensified the localized effects of light-induced renormalization. On the other hand, the light-induced renormalization effect becomes much reduced for photoexcitation with $ \lambda= 635$ nm laser even in the nanoscale-strained regions, which is consistent with the lower photoexcited QP energies that result in reduced QP funneling and interactions. 

For better understanding of our STM studies, we have further carried out first principle calculations by implementing spatially non-uniform nanoscale strain on the electronic bandstructure of TMDs through a tight-binding approach, which differ from previous studies by others that limit to spatially uniform uniaxial or biaxial strain. Our calculations further confirm the formation of strain-induced energetically favorable nanoscale regions in ML-TMDs for photoexcited QPs to funnel into. This insight holds considerable promise for the development of highly controllable optoelectronic devices, wherein tuning of the electronic bandstructure and LDOS can be achieved through controlling the energy landscape for QPs via nanoscale strain engineering, tuning the energies of QPs by varying the laser wavelength, and manipulating the densities of photoexcited QPs by varying the laser power. By leveraging on these distinctive non-equilibrium states of nanoscale strained ML-TMDs under light, there is potential for engineering TMD-based nano-optoelectronics, including photodetectors and quantum emitters with enhanced efficiency and performance. Additionally, solid-state quantum simulators for interacting exciton polaritons may be devised by engineering periodic arrays of nanoscale trapping potentials in ML-TMDs and by controlling the intensity, wavelength, and orbital angular momentum of the excitation light \cite{doi:10.1126/sciadv.abm0100}, to manipulate the density, mobility, and wavefunction of the exciton polaritons in ML-TMDs.

\section{Methods}

\subsection{Preparation of MoS$_2$-Au(111) heterostructure}

Monolayer (ML) MoS$_2$ samples were grown by chemical vapor deposition (CVD) on a SiO$_2$/Si substrate through a process detailed in Ref. \cite{Bilgin2015}, yielding single crystals on the order of a few tens of micrometer scale. Above the ML-MoS$_2$, an 80 nm layer thick Au layer was deposited via thermal evaporation without any adhesion layer. Then, epoxy was used to adhere the Au layer to a supporting glass substrate. Given that the binding energy between S atoms of MoS$_2$ and Au atoms is greater than the binding energy between S atoms and Si atoms, MoS$_2$ is transferred to Au when cleaving the Au off the Si substrate \cite{Velický2018, Wu2020, https://doi.org/10.1002/adma.201506171}.

\subsection{Scanning tunneling microscopy/spectroscopy measurements of optically excited steady states}

The LDOS of ML-MoS$_2$-Au(111) heterostructure were probed through a homemade STM head connected to a RHK R9 controller. The STM head was specially designed to have an optical window slot that allows for optical excitation from a continuous wave laser source. To ensure that excited carriers are pumped into the conduction bands, the sample was illuminated with either 2.41 eV photons ($ \lambda = 515$ nm) or 1.95 eV photons ($ \lambda = 635$ nm) generated by laser diodes. The laser was focused to a 5 mm spot size on top of the sample right underneath the STM tip with a power density up to 7 mW/cm$^2$, and with a linear polarization such that its electric field is parallel to the sample plane. The illumination of laser is performed in a continuous wave (CW) mode. Each tunneling conductance map with and without the illumination of light is performed independently. Initially, the “dark” images (without light) were captured and then the laser was turned on. Since illumination of laser inevitably causes a minor degree of local heating particularly with the tip, after turning on the laser, feedback for the tip-sample junction was kept in order to compensate for any thermal expansion. A 30-minute buffer time was accumulated while performing drift correction to maintain the tip in the original location before any measurements were made so that thermal drifts reached a steady state. Moreover, any slight drift in the constant bias conductance maps were compensated through post-measurement image processing.

All STM measurements were performed at room temperature and ambient pressure, where it has been both theoretically and experimentally demonstrated that exciton lifetimes in MoS$_2$ last on the order of nano-seconds \cite{Palummo2015, Shi2013, 10.1063/1.3636402}. Scanning tunneling spectroscopy (STS) measurements were performed using an AC lock-in amplifier with a bias modulation of $dV = 100$ mV and a drive frequency of 4 kHz to attain satisfactory signal-to-noise ratio. Furthermore, a mechanically cleaved Pt-Ir wire was used as the STM tip.

\subsection{First principle calculations}

\begin{figure}[t]
\includegraphics[width = 0.45\textwidth]{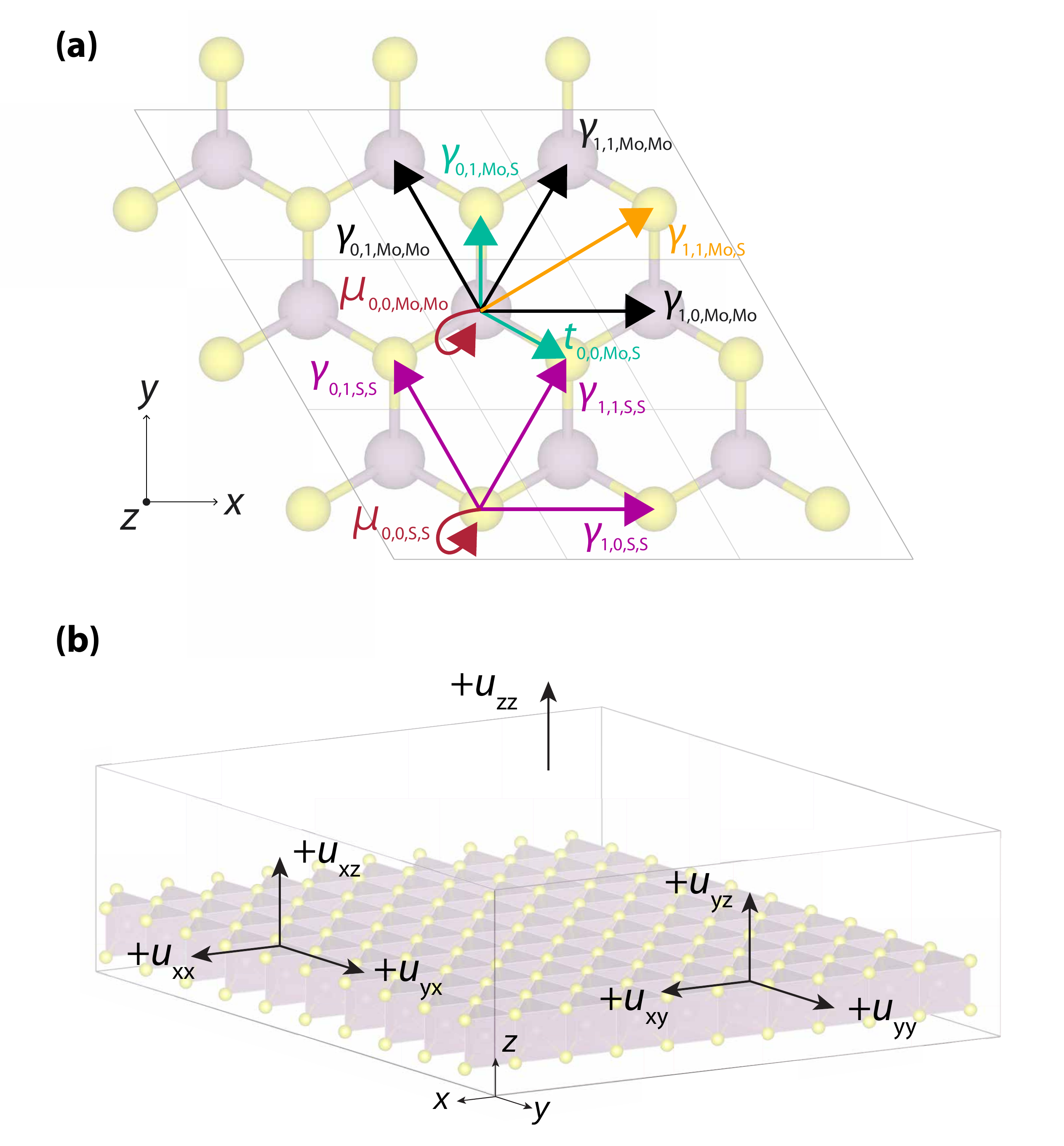}
  \caption {Hopping integrals of the tight-binding lattice Hamiltonian: (a) A schematic diagram of the lattice ab-plane of MoS$_2$ indicating the intra-unit cell (and interatomic) hopping ($\mu_{lij}$) and inter-unit cell hopping matrix element ($\gamma_(ll'ij)$) that are modified due to strain. The green arrows exhibit nearest neighbor (NN) hopping between orbitals of the Mo and S atoms, while the black and purple arrows indicate the second nearest neighbor (2NN) hopping between orbitals of Mo-Mo and S-S atoms, respectively. The orange arrow specifies the third nearest neighbor (3NN) Mo-S hopping. Intra-atomic orbital hopping is not subject to strain-induced modification. (b) A schematic diagram indicating how the atomic positions and the lattice constants are altered by the strain tensors.}
  \label{Fig11}
\end{figure}

To understand the in-plane photoexcited QP funneling into nanoscale strained regions of ML-MoS$_2$, the relationship between the strain landscape and the resulting local eigenstate modification owing to the strain is obtained through first principle calculations, which were carried out through computing the electronic states by means of density functional theory (DFT) calculations in Quantum Espresso (QE) \cite{Giannozzi_2009, Giannozzi_2017}. Here a spinless model is considered, omitting the effects of spin-orbit coupling (SOC), hence reducing the computational load and yet sufficiently capturing the effect of strain on the electronic states. The strain is expressed as a (3 $\times$ 3) strain tensor $u_{ij}$ ($i, j : \{x, y, z\}$), and is incorporated into calculations through modifying unstrained the atomic positions and the lattice vectors from $\vec{r}$ to $\vec{r'}$ through the following transformation (Fig. \ref{Fig11}(b)):

\begin{align}
\vec{r'}  & = 
\begin{pmatrix}
u_{xx} & u_{xy} & u_{xz} \\
u_{xy} & u_{yy} & u_{yz} \\
u_{xz} & u_{yz} & u_{zz} \\
\end{pmatrix} \vec{r} + \vec{r} \notag \\
& = \sum_{i=x,y,z} u_{ij} r_i \hat{e}_i + \vec{r}.
\label{strain}
\end{align}

The $x$-direction is chosen to be the same direction as the crystal $\vec{a}$-axis, and $z$ is parallel to the $c$-axis in which the diagonal elements correspond to tensile strain, and the off-diagonal matrices $u_\mathrm{ij}$= $u_\mathrm{ji}$  correspond to shear strain. The strain-modified unit cell is uniformly strained in a ($12 \times 12 \times 1$) $k$-point mesh to obtain self-consistent electron densities via self-consistent field (SCF) calculations, which are followed by non-self-consistent field (NCF) calculations. Since the experimental strain landscape induces a non-uniform strain, translational symmetry is broken, requiring change of basis from reciprocal space to real space. Bloch states are therefore used to construct maximally localized Wannier functions (MLWF) and a corresponding real-space Hamiltonian from the Wannier90 package \cite{MOSTOFI20142309}. Given that the five $d$-orbitals of the Mo atom and the six $p$-orbitals of the two S atoms in each unit cell are the only relevant bands residing close to $E_\mathrm{F}$, the unit cell Hamiltonian of this basis is a ($11 \times 11$) matrix.

As STM measurements do not directly provide information on the strain-tensor induced on the ML-MoS$_2$, STM topographic images were used to infer the corrugations in the Au(111) substrate underneath, which was then subject to molecular dynamics (MD) simulation in the large-scale atomic/molecular massively parallel simulator (LAMMPS) to gain insight on the strain of ML-MoS$_2$ that wraps over the Au corrugations. Specifically, this was done through initially setting a free standing ML-MoS$_2$ suspended 2 nm above the Au(111) substrate parallel to one another, which was then put in contact to the Au(111) substrate by moving them closer with an initial velocity of 0.005 m/s in the lateral direction. The dynamic evolution of atoms in ML-MoS$_2$ is dictated by a Stillinger-Weber (SW) potential employing the parameters outlined in Ref. \cite{Zhou_2017} , while the Au(111) substrate is regarded as a rigid body. Moreover, a Lennard-Jones potential was chosen to model the pairing interaction between Au and S of the bottom layer of MoS$_2$ directly in contact with Au substrate, which is given by
\begin{align}
U_\mathrm{Au-S} &= -4\varepsilon_\mathrm{Au-S} \Bigg[ \bigg( \frac{\sigma}{r} \bigg)^{12} - \bigg(\frac{\sigma}{r} \bigg)^{6} \Bigg].
\label{LJ}
\end{align}
Here, the interaction energy is $\varepsilon_\mathrm{Au-S}$ = 0.575 eV \cite{Tumino2020}, and the distance parameter $\sigma = (2^{1/6} \cdot d_\mathrm{Au-S}) = 2.34 $ \AA$\hspace{1 pt}$is determined using the Au-S separation of $d_\mathrm{Au-S} = 2.53$ \AA$\hspace{1 pt}$. Moreover, a cutoff distance for the Lennard-Jones potential of 3.3 \AA$\hspace{1 pt}$ was employed. After the relaxation process, from the atomic displacements resulting from the strain, the Green-Lagrangian strain tensor ($u_{ij}$) for each atom was calculated.

Since a non-uniform strain field is applied on ML-MoS$_2$, a real-space finite superlattice Hamiltonian ($\hat{H}_0$) is constructed by tiling an ($L \times L$) array of isolated and unique unit cell Hamiltonians ($\hat{H}_l$) given by the following expression:
\begin{align}
\setstackgap{L}{1.1\baselineskip}
\fixTABwidth{T}
\hat{H_0} & = 
\parenMatrixstack{
H_{1} &  \cdots & \cdots & \cdots & \cdots & H_{L}\\
\vdots & \ddots & & & & \vdots \\
\vdots &  & H_{l} & H_{l+1}& & \vdots \\
\vdots &  & H_{l+L} & H_{l+L+1}& & \vdots \\
\vdots & & & & \ddots & \vdots \\
H_{L^2-L+1} & \cdots & \cdots & \cdots & \cdots & H_{L^2}
}, 
\end{align}
where $\hat{H}_l$ is an ($11 \times 11$) block Hamiltonian matrix of the $l$-th unit cell. The onsite energies ($\varepsilon_i$) of the $i$-th orbitals and the intra-atomic hopping (e.g., Mo: $d_{xz} \rightarrow d_{yz}$) from the $i$-th to the $j$-th orbital ($\mu_{ij}$) are identical for every unit cell, as they are not affected by the compression/expansion of bond due to strain. On the other hand, inter-atomic hopping ($t_{lij}(u)$) within a unit cell is distinct for each $\hat{H}_l$, as they are influenced by strain-induced modification in bond length and bond angle. Therefore, the unit cell Hamiltonian is expressed as
\begin{subequations}
\begin{align}
\hat{H}_{l} & = \sum_{i}^{N} \varepsilon_{i} \hat{\psi}_{li}^{\dagger}\hat{\psi}_{li} \label{Honsite}\\ 
& + \sum_{\substack{\mathrm{X}_j =\mathrm{X}_i \\ j \neq i}}^{N} \mu_{ij} \cdot (\hat{\psi}_{li}^\dagger\hat{\psi}_{lj} + \hat{\psi}_{lj}^\dagger\hat{\psi}_{li}) \label{Hintraatomic} \\
& + \sum_{\substack{\mathrm{X}_j \neq \mathrm{X}_i}}^{N} t_{lij}(u) \cdot (\hat{\psi}_{li}^\dagger\hat{\psi}_{lj} + \hat{\psi}_{lj}^\dagger\hat{\psi}_{li}).
\label{Hinteratomic}
\end{align}
\end{subequations}
Here $N = 11$ is the number of orbitals per unit cell, and  $\hat{\psi}^\dagger (\hat\psi)$ is the fermionic creation (annihilation) field operator.  The first term (Eq. 5a) specifies the diagonal onsite matrix elements, while the second term (Eq. 5b) establishes the intra-atomic hopping within the atomic species $X_i \in$ [Mo, S1, S$_2$]. The last term (Eq. 5c) indicates the inter-atomic hopping between different atomic species, $X_i$ and $X_j$, which is dependent on the local strain field $u(\vec{r})$. With individual  $\hat{H}_l$  laid out to form an array, each  $\hat{H}_l$  is then stitched to form a unified lattice via nearest neighbor unit cell hopping ($\hat{H}_I$), thus allowing us to express the explicit Hamiltonian ($\hat{H}$) in the second quantization format as
\begin{align}
\hat{H} & = \hat{H}_0 + \hat{H}_\mathrm{I} \notag \\
& = \sum_l^{L^2} \bigg( \hat{H}_{l} + \hat{H}_{l,l+1} +  \hat{H}_{l,L} +  \hat{H}_{l, L + 1} \bigg), \label{latticeH}
\end{align}
in which the inter-unit cell hopping ($\gamma_{ll'ij}(u)$) contributes to the the off-diagonal term ($\hat{H}_{ll'}$), expressed as
\begin{align}
\hat{H}_{ll'} = \sum_{\substack{ij // l \neq l' }}^N \gamma_{ll'ij}(u) \cdot ( \hat{\psi}_{li}^\dagger\hat{\psi}_{l'j} + \hat{\psi}_{lj}^\dagger\hat{\psi}_{l'i} ) .
\end{align}
This ansatz leads to the breaking of C$_3$ symmetry due to strain, resulting in an increased set of inter-atomic hopping parameters outlined in Fig. \ref{Fig11}(a). There are three categories of inter-atomic hopping: (1) nearest neighbor (NN) hopping between Mo-S orbitals, (2) second-nearest neighbor (2NN) hopping between Mo-Mo or S-S orbitals, and (3) third-nearest neighbor (3NN) hopping between Mo-S \cite{PhysRevB.92.205108, PhysRevB.98.075106}. These hopping parameters are derived by applying the local strain configuration obtained from the MD simulations to a single unit cell with periodic boundary conditions, thereby inducing a uniform deformation. Subsequently, independent high-throughput DFT calculations are conducted for the differently deformed unit cells, yielding the hopping parameters unique to a specific strain. For a finite lattice with size $L = 55$, the Hamiltonian involves a total of $55^2 = 3025$ independent DFT calculations for each unit cell strain. To mitigate the substantial computational costs, strain tensors were categorized into a reduced set of clusters. Calculations were then carried out for each cluster, offering a more efficient approach to manage the computational expense.


\section{Author Contributions}
A.P. and N.-C.Y. conceived the research ideas. A.P. constructed the optical-STM system, and carried out all measurements and data analyses together with R.K.. A.P. contributed to all DFT and FDTD calculations. A.P., R.K. and Y.C. contributed to the MD simulations. J.G. carried out the growth of MoS$_2$ samples by CVD and characterized them by PL and Raman measurements. D.H. contributed to evaporation of Au and the transfer of samples. A.P. wrote the first draft of the manuscript with contributions from R.K.. N.-C.Y. edited the manuscript, and finalized the manuscript with A.P.. All authors read and approved the final manuscript. N.-C.Y. supervised and coordinated the project.

\section{Acknowledgement}
This work was primarily supported by the National Science Foundation under the Physics Frontier Centers program for the Institute for Quantum Information and Matter (IQIM) at the California Institute of Technology (Award $\#$1733907). N.-C.Y. acknowledges partial support from the Thomas W. Hogan Professorship at Caltech, the Yushan Fellowship awarded by the Ministry of Education in Taiwan, and the Yushan Fellow Distinguished Professorship at the National Taiwan Normal University in Taiwan. J.G. acknowledges support from BaCaTeC (Project No. 14 [2022-2]), the DFG via SPP2244 (Project ID 443405595), and the Munich Center for Quantum Science and Technology (MCQST) (EXC 2111, Project ID 390814868).

\putbib

\defaultbibliographystyle{apsrev4-2}
\defaultbibliography{apssamp}
\end{bibunit}

\newpage   

 \title{Supporting Information: \\Strongly Enhanced Electronic Bandstructure Renormalization by Light \\ in Nanoscale Strained Regions of Monolayer MoS$_2$/Au(111) Heterostructures}%
 \vspace{300 pt}
 \maketitle

\FloatBarrier
\newcommand{\beginsupplement}{%
        \setcounter{table}{0}
        \renewcommand{\thetable}{S\arabic{table}}%
        \setcounter{figure}{0}
        \renewcommand{\thefigure}{S\arabic{figure}}%
        \setcounter{equation}{0}
        \renewcommand{\theequation}{S\arabic{equation}}%
     }  
\beginsupplement

\begin{bibunit}
\setcounter{page}{1}

\onecolumngrid

\section{Supplementary Information Note 1\\Photoluminescence and Raman spectra of monolayer (ML) MoS$_2$}
\begin{figure}[h!]
\includegraphics[width = 0.4\textwidth]{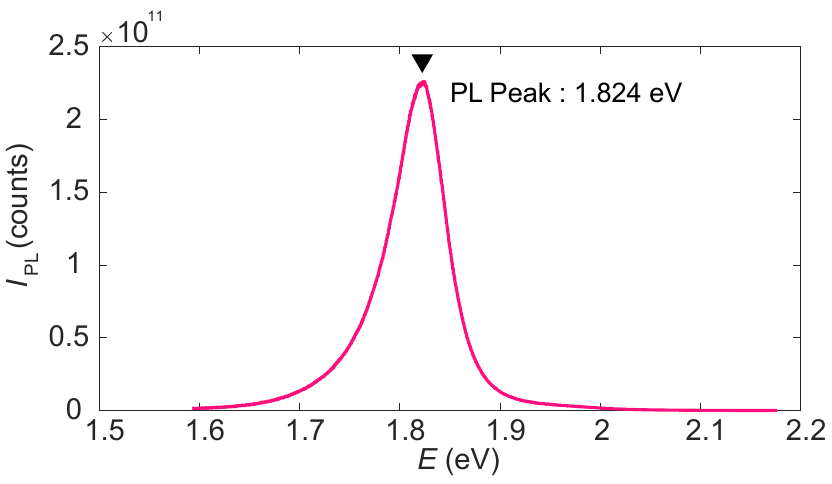}
  \caption {PL spectrum of ML-MoS$_2$ on a SiO2 / Si substrate, measured at room temperature.}
  \label{FigS1}
\end{figure}

\begin{figure}[h!]
\includegraphics[width = 0.4\textwidth]{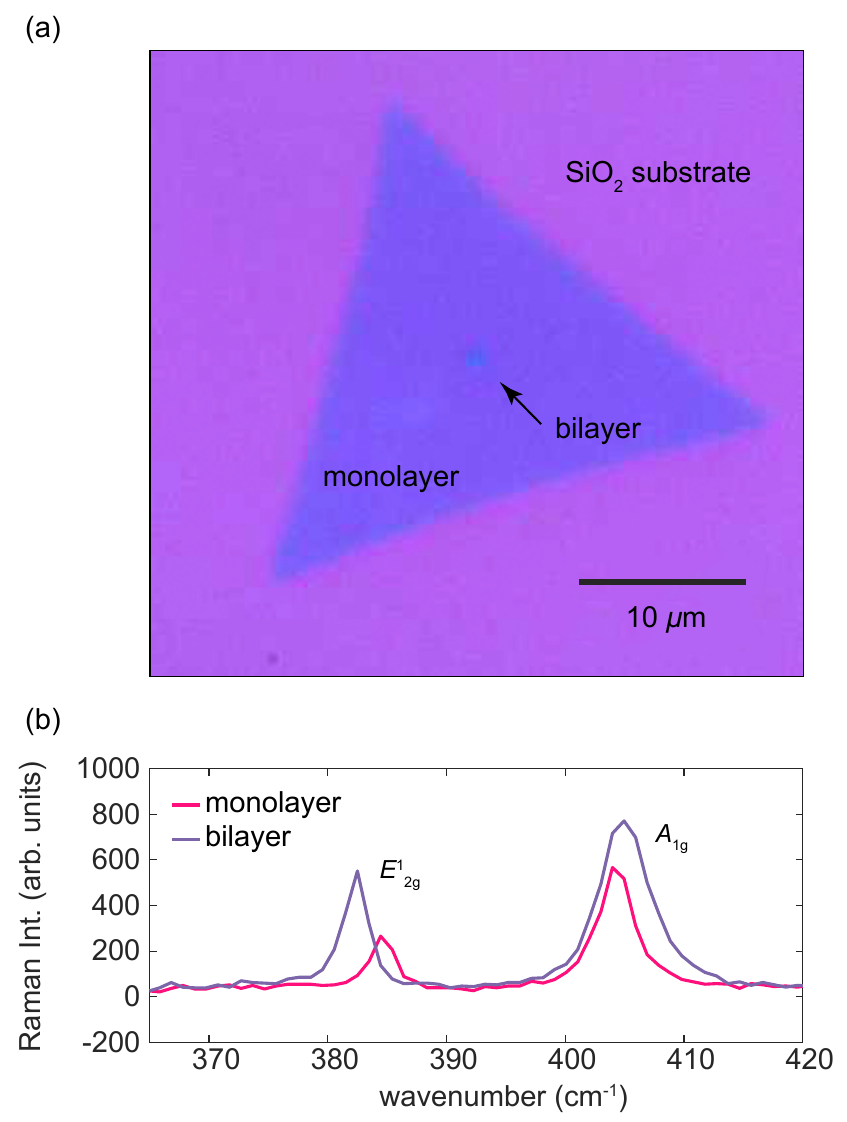}
  \caption {(a) An optical microscope image of MoS$_2$ with both a monolayer and bilayer region on a SiO$_2$ / Si substrate. (b) Raman spectra of the monolayer and bilayer region measured at room temperature.}
  \label{FigS2}
\end{figure}

The photoluminescence (PL) and Raman spectra measured on as-grown ML-MoS$_2$ on a SiO2 / Si wafer before transferring to the Au(111) substrate are shown in Figs. \ref{FigS1} and \ref{FigS2}, respectively. For the PL spectroscopy (Fig. \ref{FigS1}), an excitation wavelength of $\lambda = 514$ nm with a power of 9 $\mu$W is used, and the signal was integrated over 20 s to obtain the spectrum. The spectrum shows a PL peak at 1.824 eV and a FWHM of 63.2 meV.

Raman spectroscopy was performed through excitation by a $\lambda = 514$ nm laser with a power of 100 $\mu$W using a laser spot size of 1 $\mu$m$^2$. The $E_{2g}^1$ and $A_{1g}$ peaks are shown in Fig. \ref{FigS2}(b) for the monolayer region and the bilayer region exhibited in Fig. \ref{FigS2}(a). The difference in the wavenumber between the two Raman modes ($\Delta \omega$) are 19.7 cm$^{-1}$ and 22.6 cm$^{-1}$, being consistent to those identified to the monolayer and the bilayer regions identified in Refs. \cite{Bilgin2015, Lee2010}.

\newpage   
\section{Supplementary Information Note 2 \\ Green-Lagrange strain tensor}

\begin{figure}[h!]
\includegraphics[width = 0.8\textwidth]{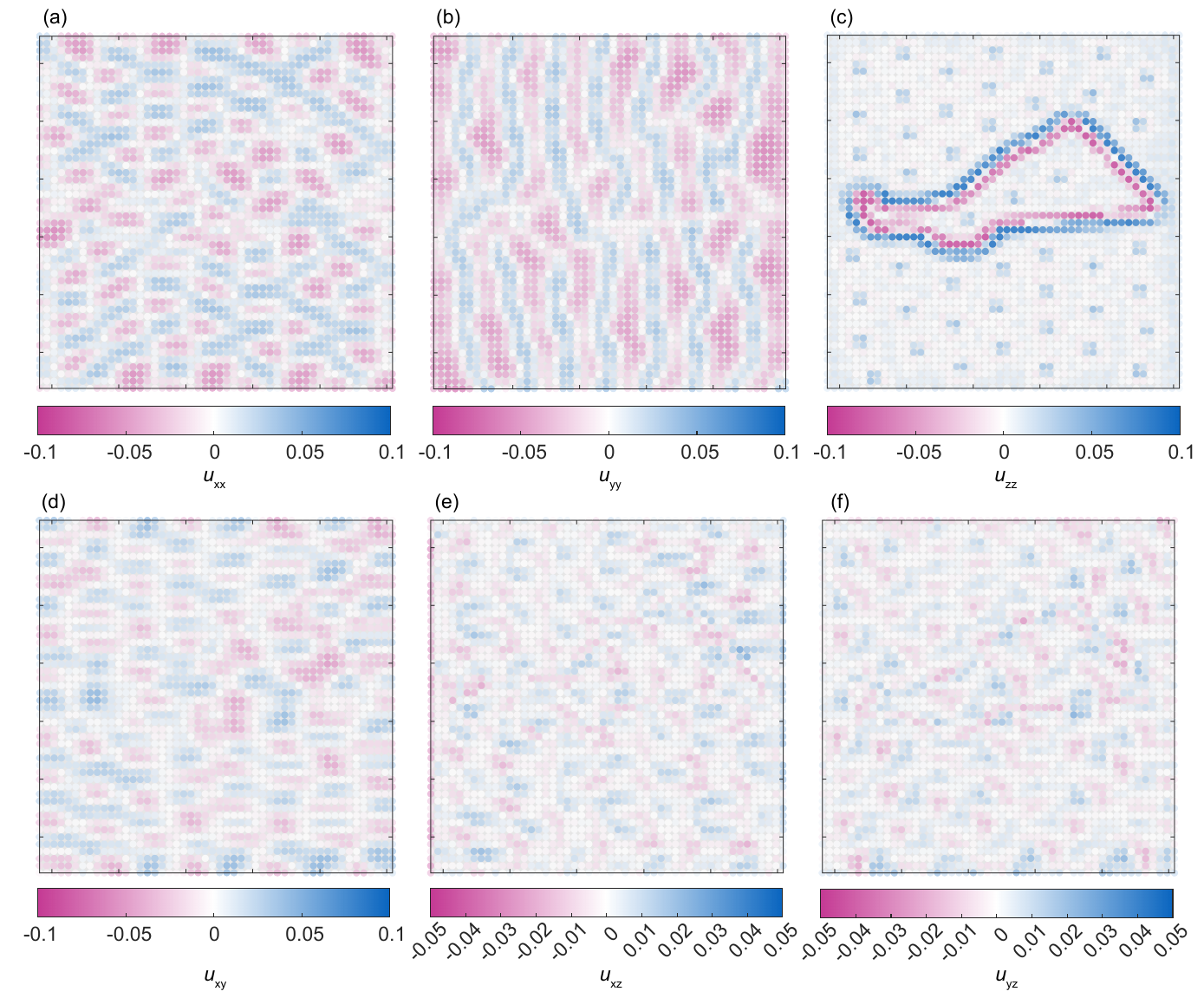}
  \caption {The Green-Lagrange strain tensor components (a) $u_{xx}$, (b) $u_{yy}$, (c) $u_{zz}$, (d) $u_{xy}$, (e) $u_{xz}$, and (f) $u_{yz}$, obtained from molecular dynamics (MD) simulations of ML-MoS$_2$ on an indented Au(111) substrate, as shown in Fig. \ref{Fig4}(d). All panels share the same color bar.}
  \label{FigS3}
\end{figure}

The strain tensor of the relaxed ML-MoS$_2$ on Au(111) substrate was calculated from a method detailed in Ref.\cite{2007MJ200769}, using the OVITO package \cite{Stukowski_2010}. A freestanding ML-MoS$_2$ before interacting with the Au substrate is used as a reference to quantify the displacement, and the cutoff radius of 8 \AA$\hspace{1 pt}$ is employed to determine the neighboring atomic sites to compute the deformation gradient potential. Figure \ref{FigS3} exhibits the unit-cell strain tensor obtained from the Au(111) indentation configuration of Fig. \ref{Fig4}(d). Here, each marker represents each a single unit cell.

\newpage   
\section{Supplementary Information Note 3 \\ Large area topographic image of MoS$_2$ wrapped over corrugated Au(111) substrate}

\begin{figure}[h!]
\includegraphics[width = 0.8\textwidth]{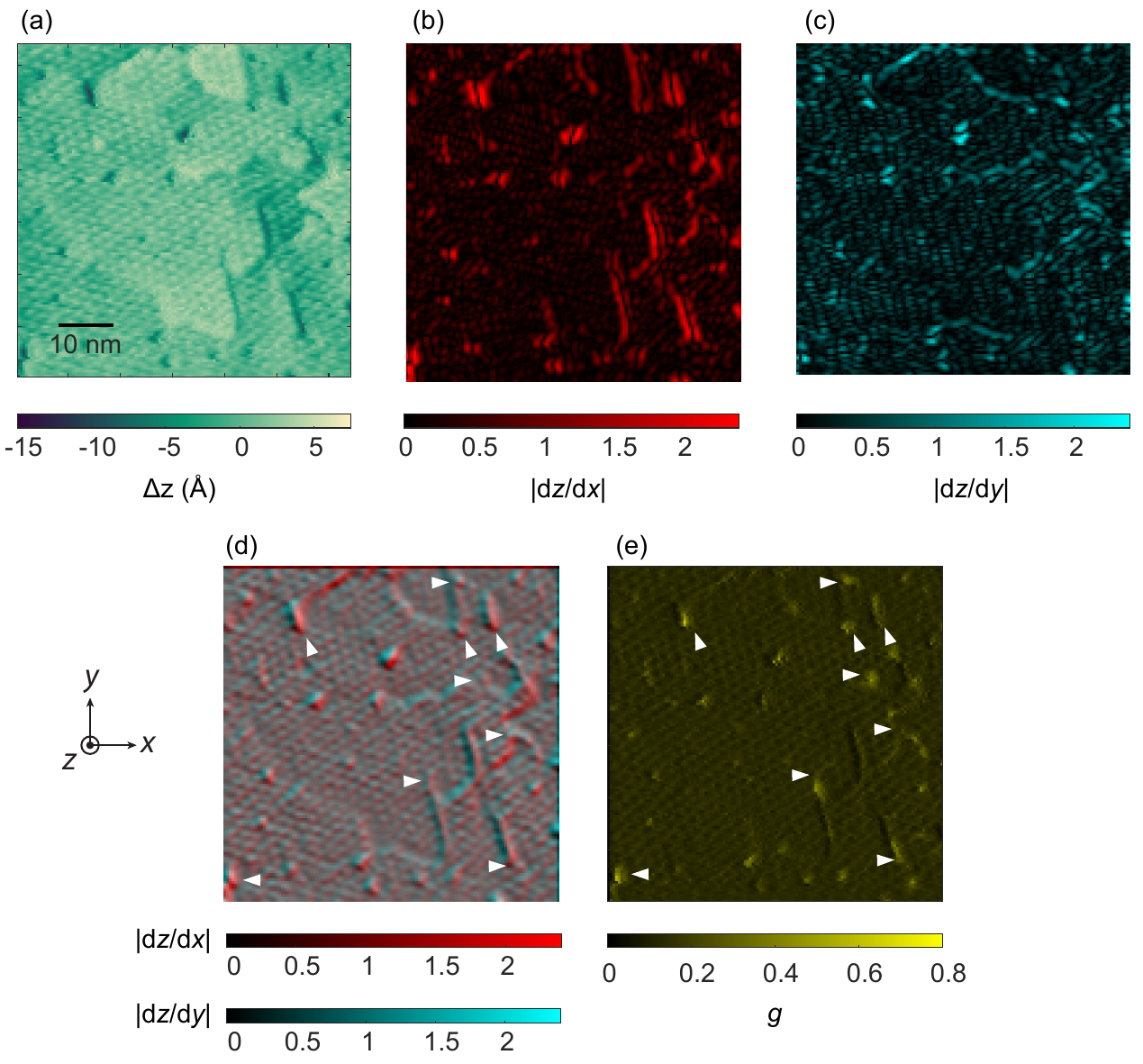}
  \caption {(a) Topographic map over a 60 nm area. A topographic gradient map in two orthogonal directions: (b) $|dz/dx|$ and (c) $|dz/dy|$. The gradient only reaches a value of $\approx 2.4$ \AA, suggestive that the indentation is only a single atomic layer thick depression in the Au(111) substrate in this area. (d) A superimposed image of the $|dz/dx|$ and $|dz/dy|$ map. (e) A conductance map measured at $V_\mathrm{b} = 600$ mV. The white arrows indicate corner-like features of indentations that exhibit particularly large LDOS.}
  \label{FigS4}
\end{figure}

By taking the gradient of a topographic map as shown in Fig. \ref{FigS4}(a-d), and comparing it with the constant bias conductance map (Fig. \ref{FigS4}(e)), it becomes clear that there are regions of high LDOS in regions that exhibit both relatively large $|dz/dx|$ and $|dz/dy|$ in the topography, resulting from a corner-like region of an indentation in the underlying Au. This phenomenon indicates that a corner-like feature in the Au indented region generally results in a stronger strain-modified LDOS.

\begin{figure}[h!]
\includegraphics[width = \textwidth]{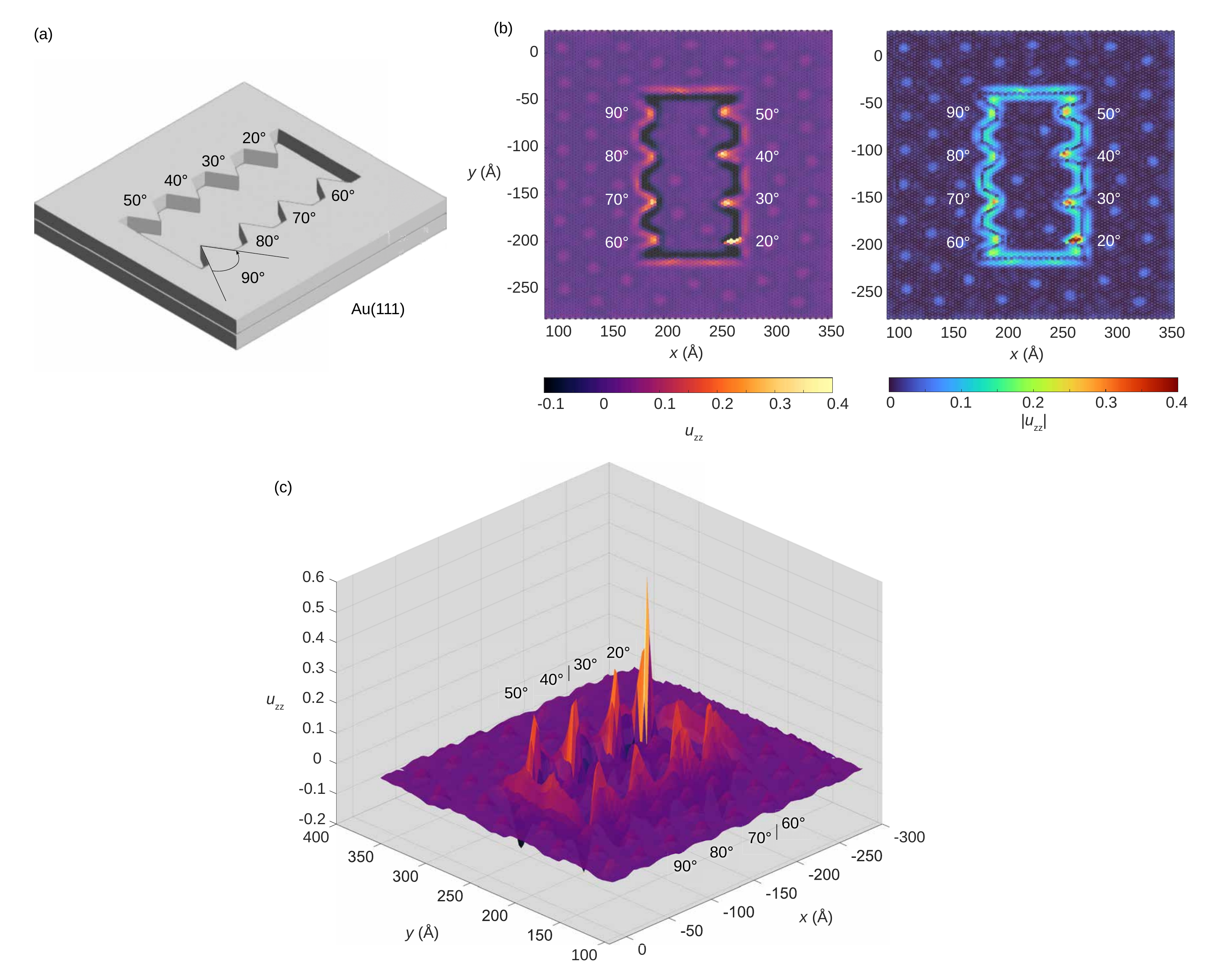}
  \caption {An Au(111) substrate with various corner-like indentations of different angles subject to MD simulations. (b) The $u_{zz}$ strain tensor and its magnitude $|u_{zz}|$ of MoS$_2$ relaxed on Au(111) substrate with an indentation displayed in panel (a). (c) A 3D surface plot depicting the spatial distribution of $u_{zz}$, suggesting that the sharper corner-like regions are more highly strained.}
  \label{FigS5}
\end{figure}

The effects of corner-like structures were further investigated \textit{in-silico}. A single layer of atomic indentation in Au(111) with various corner-like features of various angles were constructed as shown in Fig. \ref{FigS5}(a). From MD simulations, a single layer of MoS$_2$ was relaxed on top, yielding a strain component $u_{zz}$ distribution depicted in Fig. \ref{FigS5}(b,c).

\begin{figure}[h!]
\includegraphics[width = \textwidth]{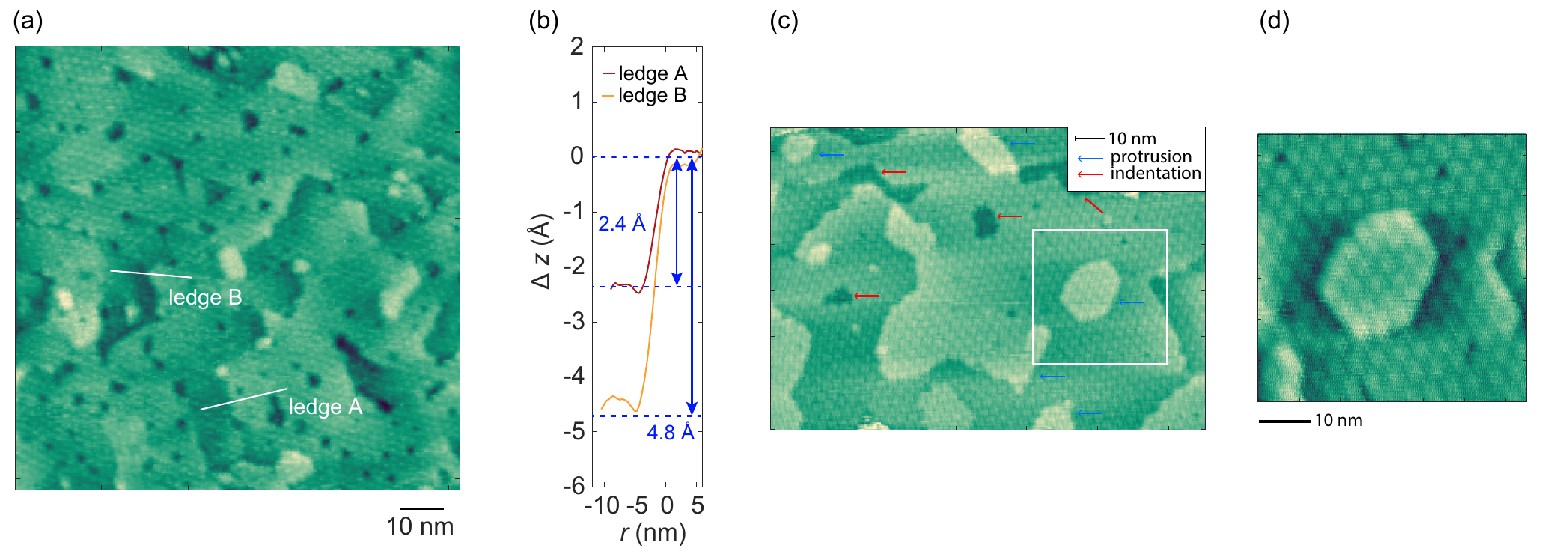}
  \caption {(a) A large scale (100 × 100) nm$^2$ topographic STM image and (b) line profiles taken along white lines represented in panel (a). (c,d) A higher resolution topographic image indicating both protrusion-type and indentation-type topographic features.}
  \label{FigS6}
\end{figure}

Figure \ref{FigS6}(a) displays a topographic map over a larger area of (100 nm $\times$ 100 nm), showing nanoscale corrugations and Moir\'e lattices. By taking a line profile along randomly chosen exemplary ledges in this region, it is evident that the height differences across the ledge are all integer multiples of the interlayer spacing of Au(111), indicating that corrugations in the STM topographic images are generally a consequence of the Au(111) terrace/indentation (Fig. \ref{FigS6}(b)).

\begin{figure}[h!]
\includegraphics[width = \textwidth]{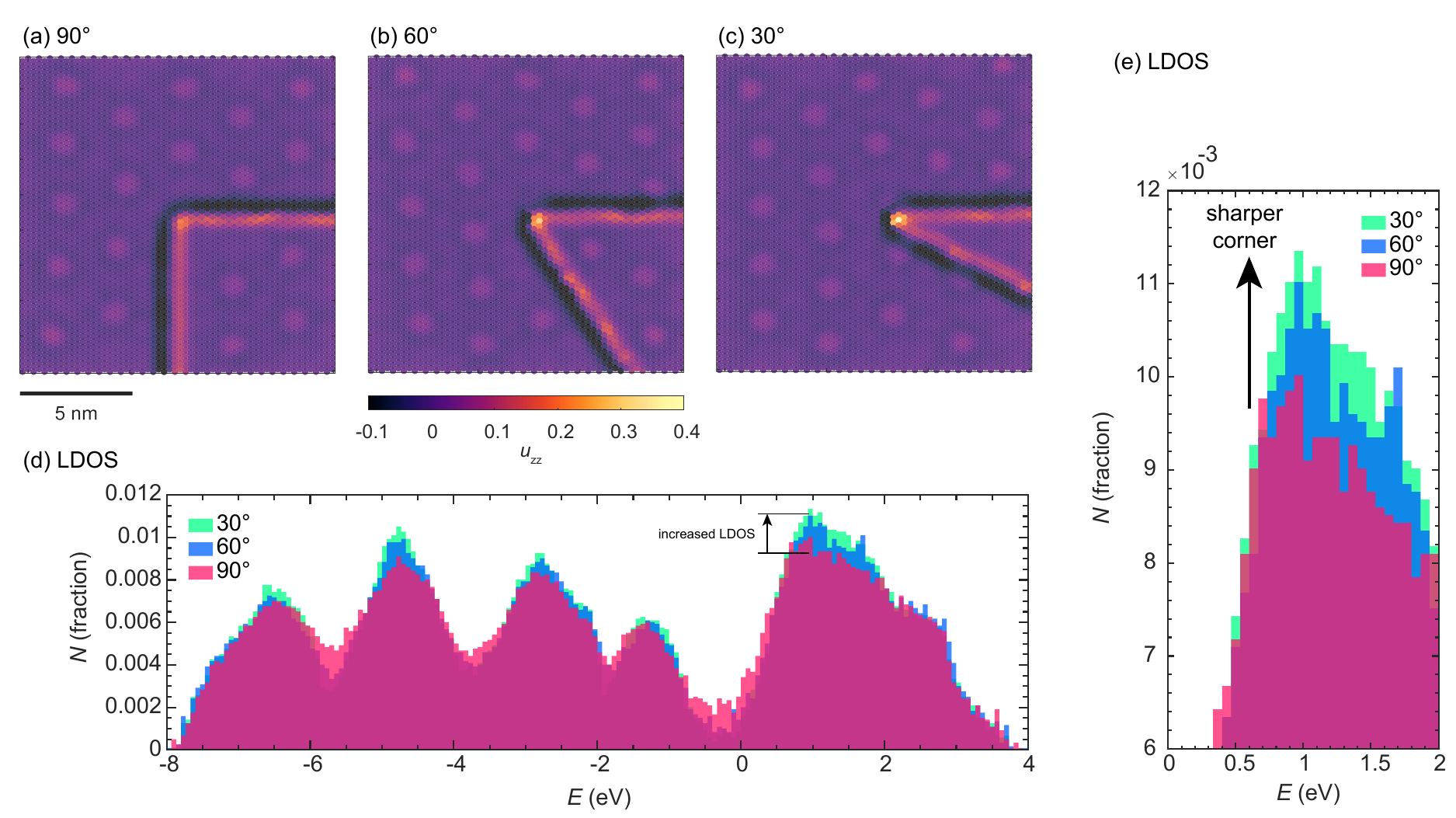}
  \caption {A $u_{zz}$-strain component distribution map for corner-like features with an angle of (a) 90$^\circ$, (b) 60$^\circ$, and (c) 30$^\circ$. (d) The eigenvalue histogram of the tight-binding Hamiltonian of strain field tensors represented in panels (a)-(c). (e) A zoomed in eigenvalue histogram shown in panel (d) of energies above the conduction band, highlighting stronger modifications associated with sharper edges.}
  \label{FigS7}
\end{figure}

The strain is pronounced at the corner-like features compared to the straight step-edges. As the corner-like feature becomes sharper, the $u_{zz}$ strain component intensifies. Specifically, an extreme strain of $u_{zz} \approx 0.3$ is observed at a corner feature with an angle of 20$^\circ$. To delve deeper into the effect of the strain on the electronic structure, a section of the corner-like feature was extracted \ref{FigS7}(a-c), and the corresponding strain tensor field was implemented onto a tight-binding Hamiltonian in which the hopping integral was adjusted in accordance to the strain. The eigenvalue spectra, representing the LDOS of the corner region, indicate that the LDOS is enhanced with sharper corner-like features among the conduction bands, which is consistent to STM measurements (Fig. \ref{FigS4}).

\newpage
\section{Supplementary Information Note 4 \\ Moir\'e lattice of multilayer MoS$_2$ on Au(111)}

\begin{figure}[h!]
\includegraphics[width = 0.5\textwidth]{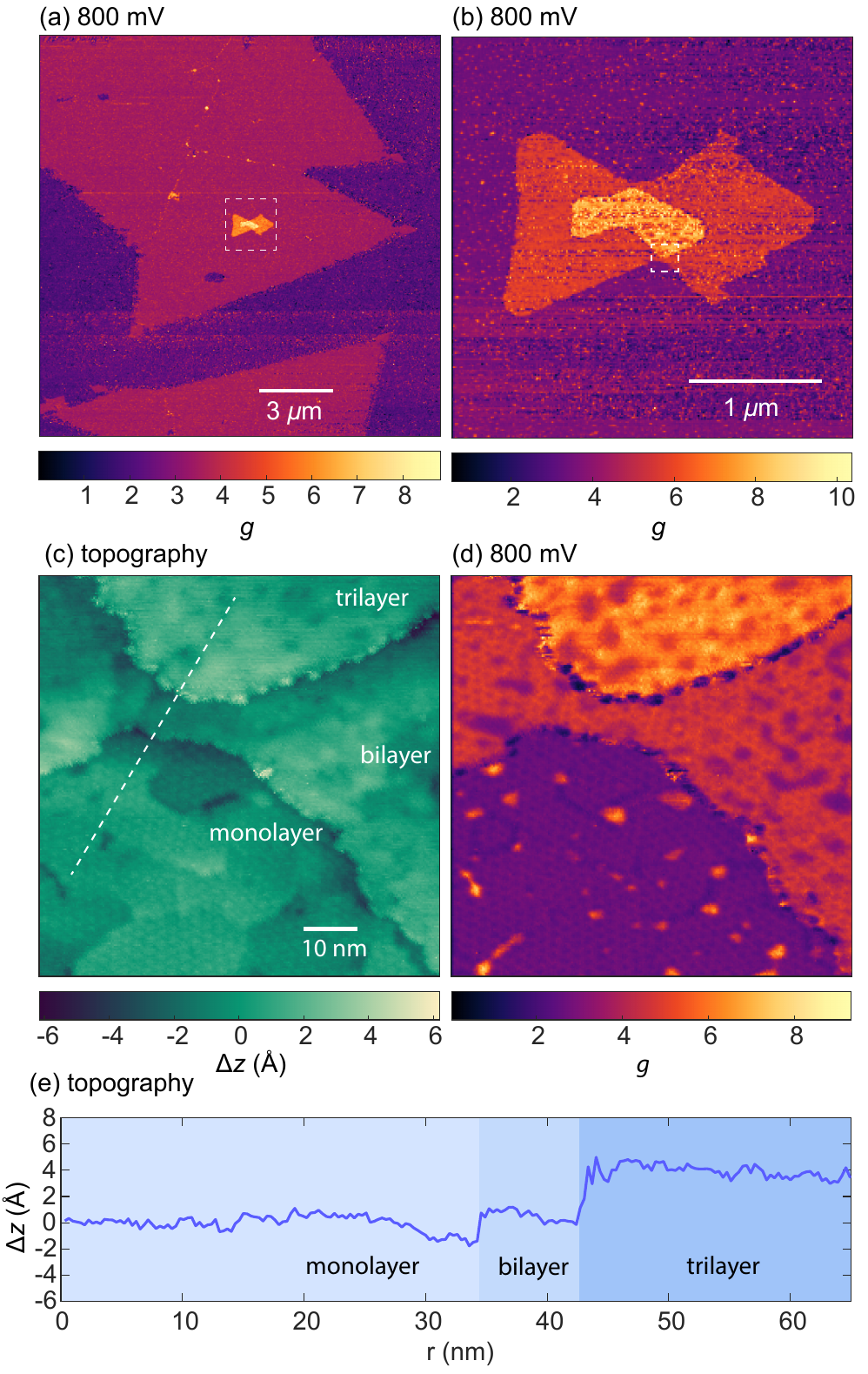}
  \caption { (a) A constant bias conductance map of MoS$_2$ layers overlapping each other on top of a larger scale monolayer, measured with a tunneling resistance of 1 G$\Omega$. (b) A zoomed-in constant bias map of the area enclosed by the white dashed box shown in panel (a). (c) A topographic map of the region enclosed by the area indicated by the white box, showing regions of monolayer, bilayer, and trilayer configurations, with (d) the corresponding conductance map at a constant bias voltage of 800 mV. (e) A line profile cut of the topographic map indicated by the white lines of panel (c).}
  \label{FigS8}
\end{figure}

Through scanning over large micron scale areas, regions exhibiting multilayer MoS$_2$ were occasionally found. On some of the monolayer triangular crystals of $\approx$ 10 $\mu$m, smaller $\approx$ 1 $\mu$m triangular monolayer crystals can be identified. The two smaller crystals overlap to create both a bilayer region and a trilayer region. The three regions of different layer thickness indicate contrasting conductance, owing to their different electronic structure as depicted in Fig. \ref{FigS8}(a,b,d). As shown in Fig. \ref{FigS8}(b), while the monolayer region indicates a clear triangular Moir\'e lattice, the Moir\'e lattice becomes faint in the bilayer region. Moreover, the Moir\'e lattice seem to merge to form irregular patterns. In the trilayer region, the Moir\'e lattice is no longer visible. A similar multilayer structure has also been observed in Ref.1, which employs a similar CVD growth process.

\newpage
\section{Supplementary Information Note 5 \\ STM tip-induced plasmonic enhancement}

\begin{figure}[h!]
\includegraphics[width = 0.8\textwidth]{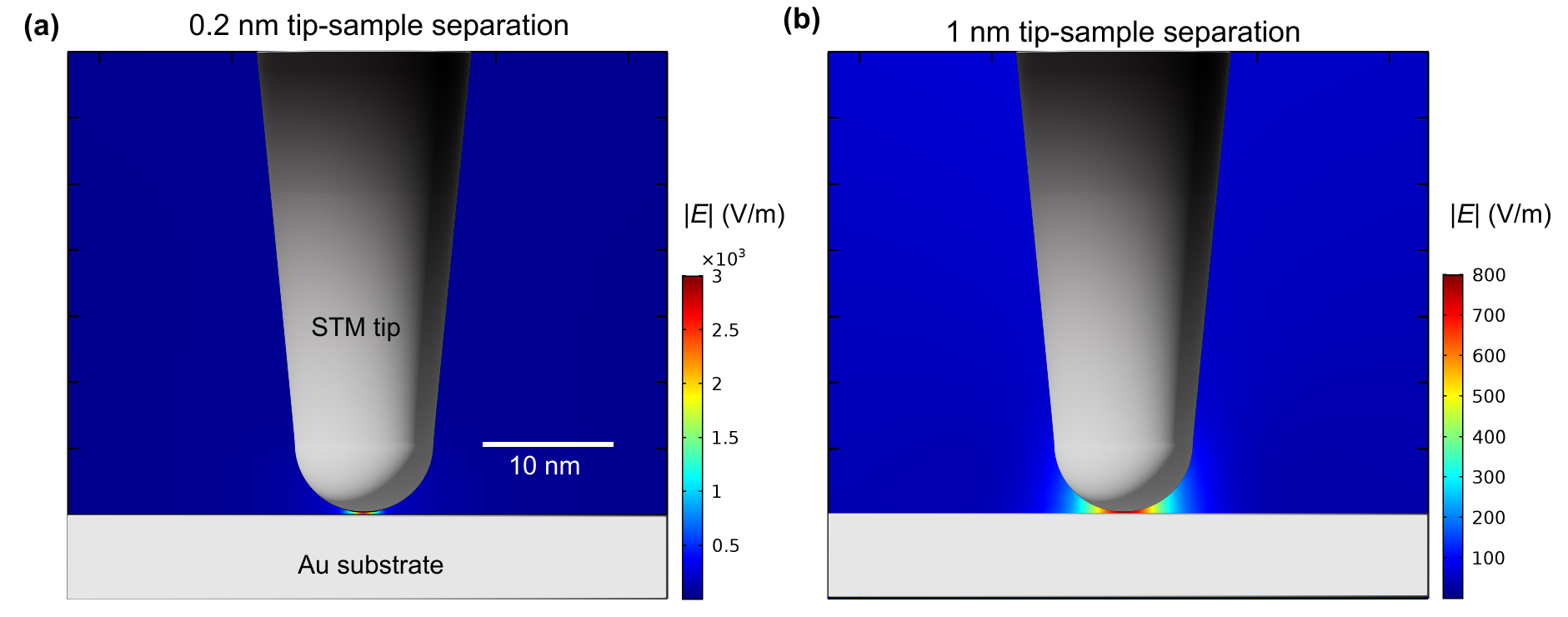}
  \caption { Calculated electric-field profiles in the tip-sample junction with a tip-sample separation of (a) 0.2 nm and (b) 1 nm.}
  \label{FigS9}
\end{figure}

Upon illuminating the tunneling junction with a laser, where the junction of STM tip/tunneling gap plus ML-MoS$_2$/Au substrate forms a metal/dielectric/metal (MDM) heterostructure, electromagnetic excitations at the interface are evanescently confined to create surface plasmon polaritons (SPPs). To simulate the electric field distribution across the tunneling junction, Maxwell’s equations were solved through a Finite-Difference Time-Domain (FDTD) approach using COMSOL Multiphysics. The tip-sample junction was modelled as depicted in Fig. \ref{FigS9}, in which the rotational invariance of the STM tip allows us to reduce the simulation geometry down to 2D.  Although the STM tip is fabricated from an Pt/Ir (1:9) alloy wire, the apex of the tip consists of Au atoms because it is standard practice in STM studies to dip the STM-tip about 1 $-$ 2 nm into an Au substrate before measurements. Given that Au is relatively soft, this procedure allows Au clusters to attach to the STM tip, thereby creating an atomically sharp tip.) Hence, we modelled the apex of the STM tip to be a 5 nm round gold ball, as shown in Fig. \ref{FigS9}, and placed it at $\approx$ 1 nm above an Au substrate, which was a typical tunneling range. We further applied a 582 THz frequency electric field (515 nm) with power of 0.7 mW/cm$^2$ from above, while setting the left and right boundaries as scattering boundaries that were transparent to scattered waves with no reflection. Plotting the norm of the electric field, we find that there is indeed a $\approx$ 10$^3$-fold enhancement in the magnitude of electric field confined within the tunneling junction, suggesting that our experimental setup can substantially increase the light-matter interactions, magnifying the exciton-induced effects with a relatively low laser power.

\newpage
\section{Supplementary Information Note 6 \\ Calculated strain-induced changes in the LDOS}

\begin{figure}[h!]
\includegraphics[width = \textwidth]{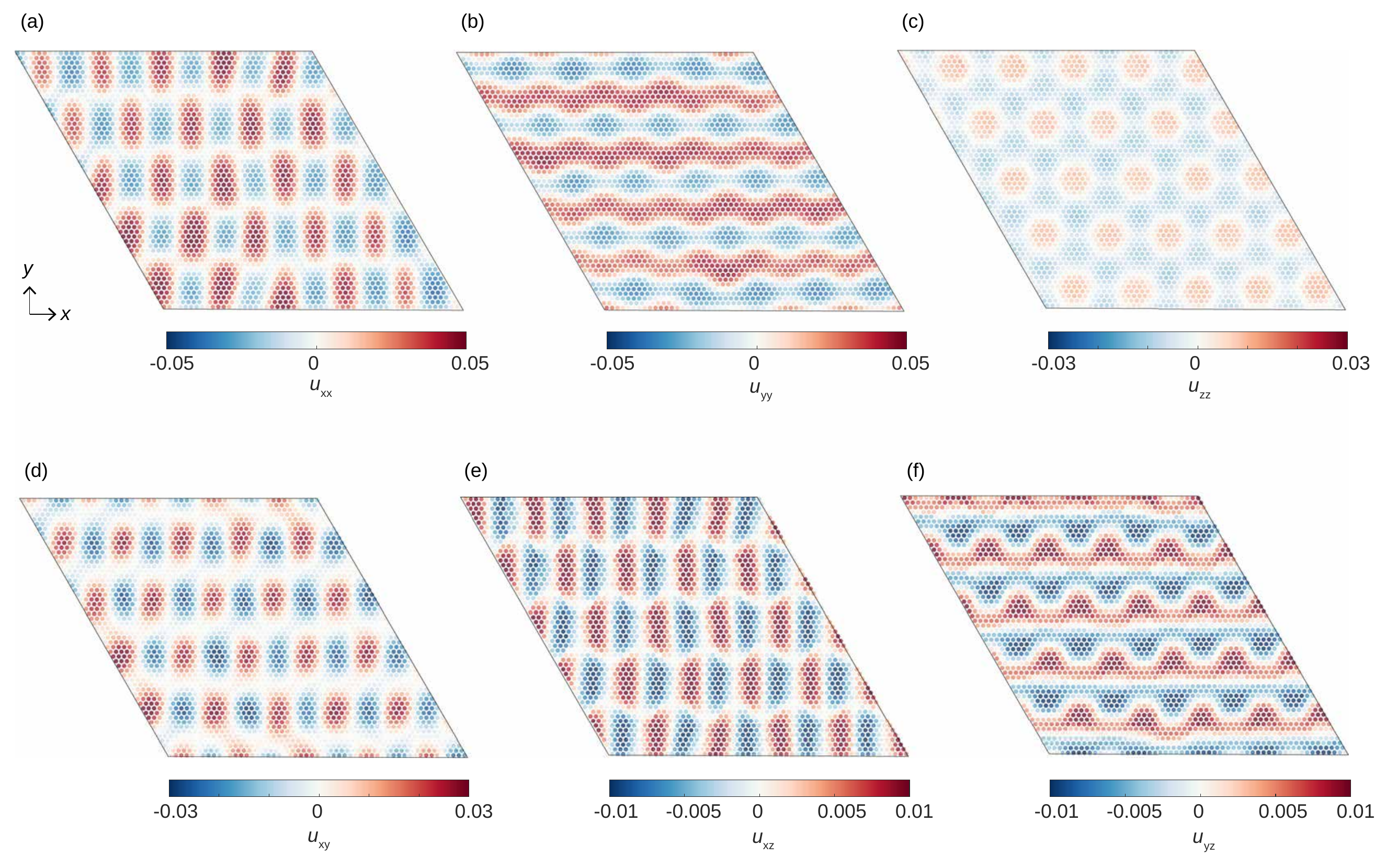}
  \caption {The Green-Lagrange strain tensor components obtained from molecular dynamics (MD) simulations of ML-MoS$_2$ on a flat Au(111) substrate, as shown in Fig. \ref{Fig7}(a).}
  \label{FigS10}
\end{figure}

\begin{figure}[h!]
\includegraphics[width = \textwidth]{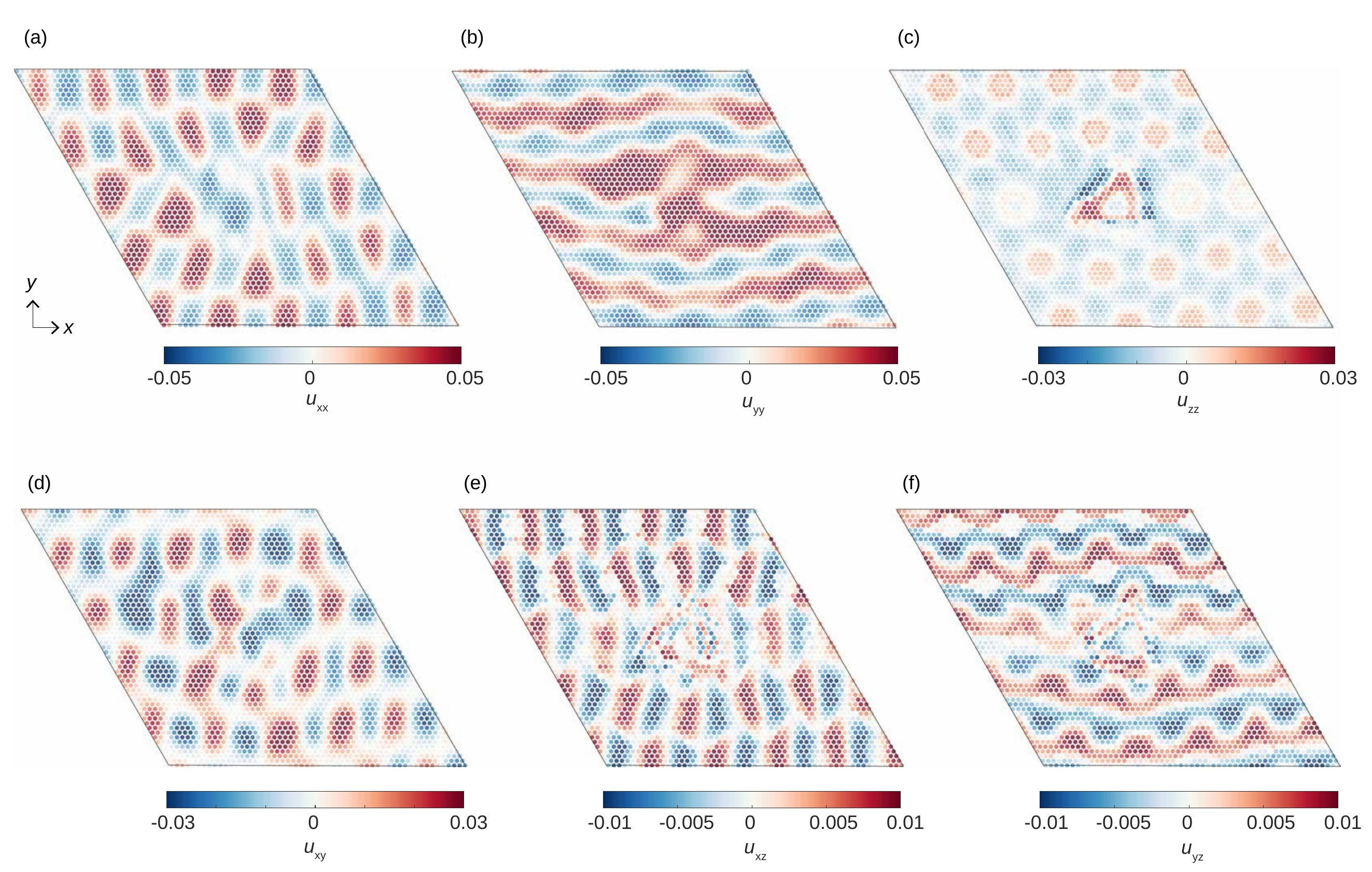}
  \caption {The Green-Lagrange strain tensor components obtained from MD simulations of ML-MoS$_2$ on an indented Au(111) substrate, as shown in Fig. \ref{Fig7}(b).}
  \label{FigS11}
\end{figure}

Fig. \ref{FigS10} and Fig. \ref{FigS11} shows the strain-field laid over each unit cell of a (55 $\times$ 55)-lattice Hamiltonian for a flat Au(111) substrate and an Au(111) substrate with a triangle indentation modelled after the indentation of Fig. \ref{Fig6}(a).

\newpage

\begin{figure}[h!]
\includegraphics[width = 0.6\textwidth]{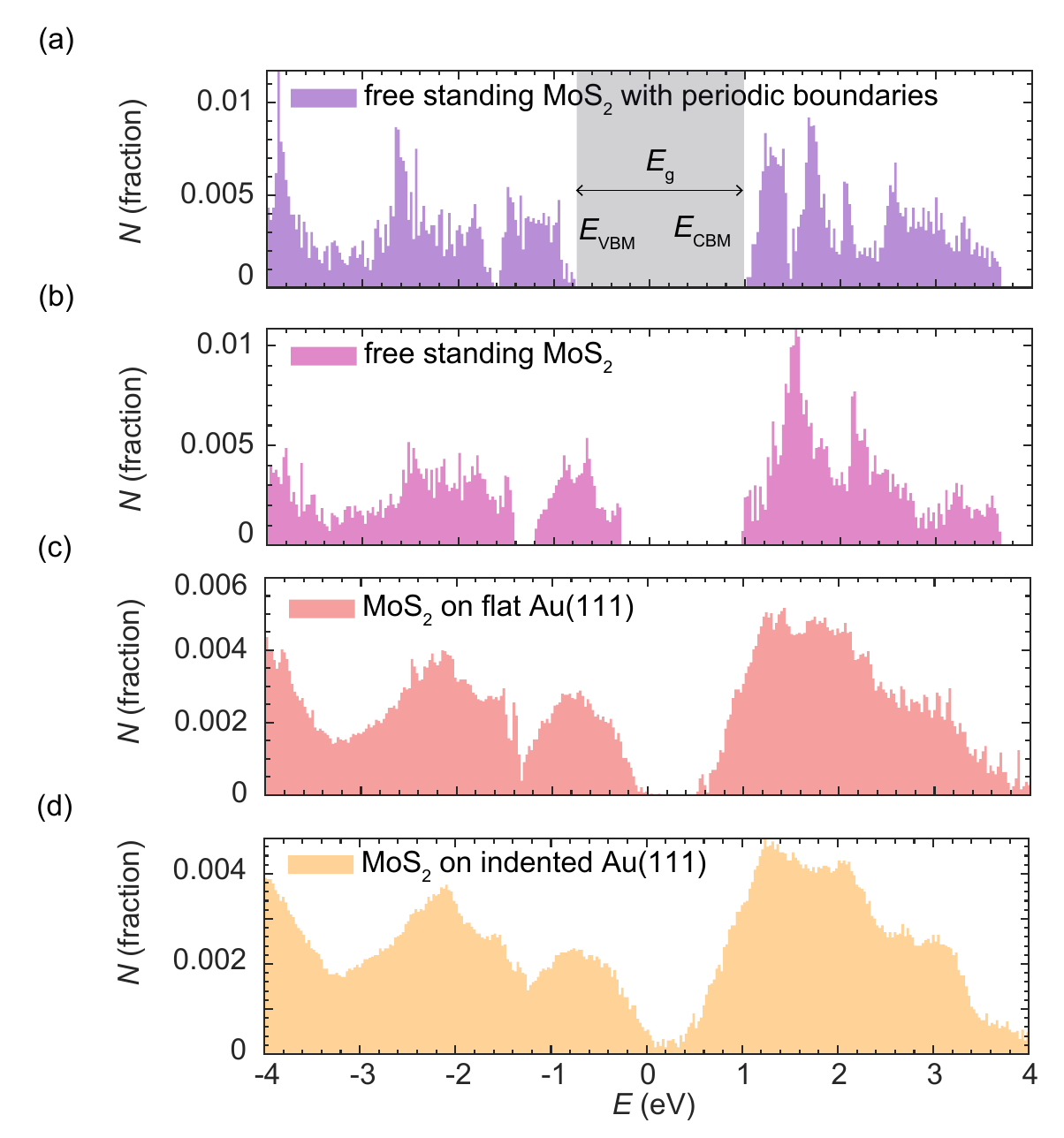}
  \caption {Energy spectra of the global DOS of the entire (55 $\times$ 55) lattice for (a) free-standing ML-MoS$_2$ with periodic boundary conditions, (b) free-standing ML-MoS$_2$ with finite boundaries, (c) ML-MoS$_2$ is laid on flat Au(111), and (d) MoS$_2$ is laid on indented Au(111). The shaded grey regions indicate the region of the band gap ($E_\mathrm{g}$) of the infinite free standing ML-MoS$_2$ lattice.}
  \label{FigS12}
\end{figure}

Cross-referencing the DOS spectrum with the calculated band structure shown in Fig. \ref{Fig3}(b), the two peak-like structures close to $E_\mathrm{F}$, is assigned as the $E_\mathrm{VBM}$ and the $E_\mathrm{CBM}$, thus exhibiting a band gap of $E_\mathrm{g} \approx 1.8$ eV. Intriguingly, for the finite lattice without periodic boundary conditions, the band gap is accompanied by sparse mid-gap states as shown in Fig. \ref{Fig7}(b) close to the valence band, which can arise due to interference and localization of eigenstates at the edge of the finite lattice due to the abrupt termination. These states within the gap are localized at the edge of the sample (Fig. \ref{Fig7}(g)), consistent with observations on MoS$_2$ nanoflakes \cite{PhysRevLett.87.196803, Grønborg2018}. Upon interfacing the ML-MoS$_2$ to the Au(111) surface, a general blurring of the features in the DOS are observed (Fig. \ref{FigS12}(c,d)).

\newpage

\newpage
\section{Supplementary Information Note 7 \\ Effect of uniform strain on the electronic bandstructure}

\begin{figure}[h!]
\includegraphics[width = 0.75\textwidth]{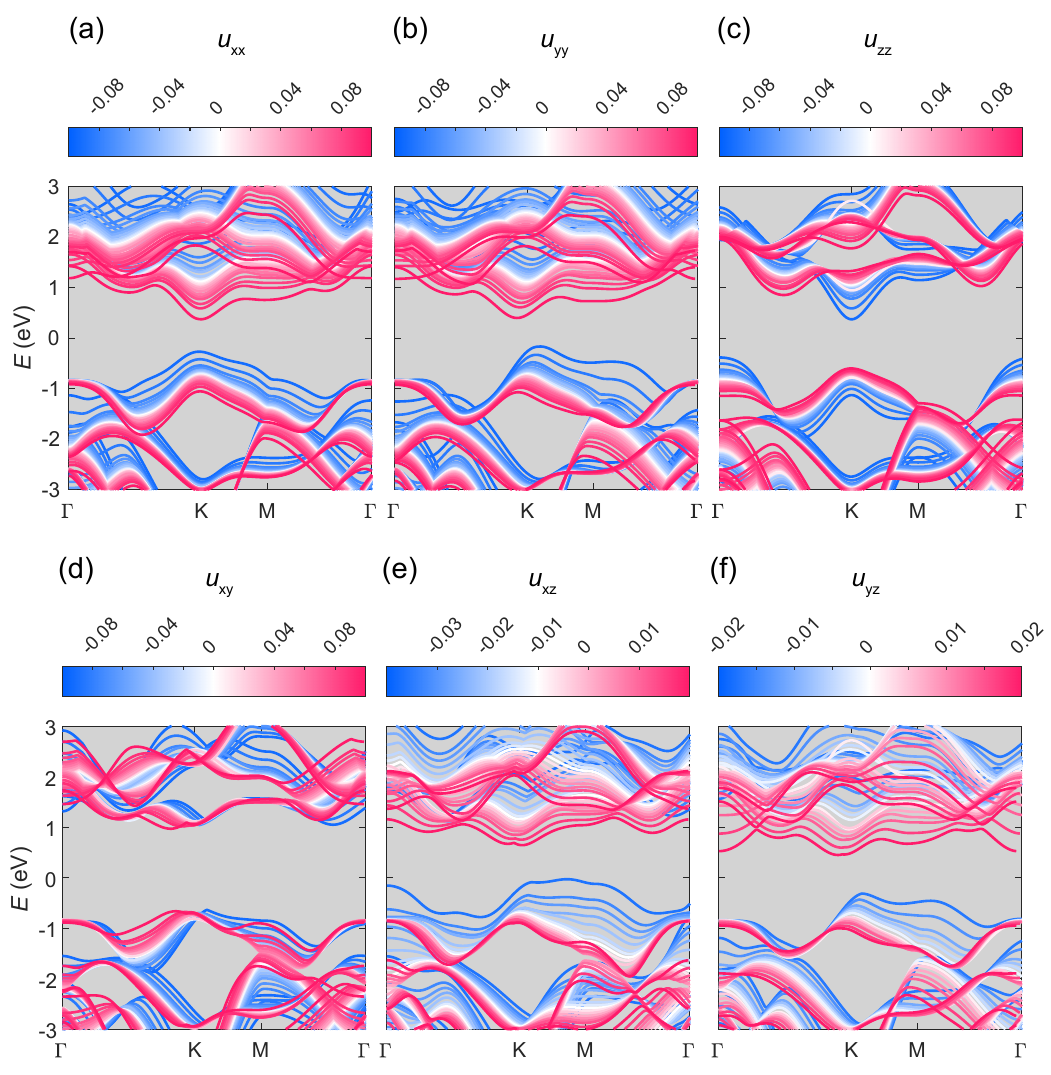}
  \caption {The strain-dependent bandstructure of MoS$_2$ assuming a uniform strain: (a) $u_{xx}$, (b) $u_{yy}$, (c) $u_{zz}$, (d) $u_{xy}$, (e) $u_{xy}$ and (f) $u_{yz}$. Positive strain corresponds to expansion while a negative strain corresponds to compression.}
  \label{FigS13}
\end{figure}

The hopping matrix elements for each strain configuration were implemented into a unit cell Hamiltonian with periodic boundary conditions. The Hamiltonian was then diagonalized to obtain the bandstructure for different strain configurations, as displayed in Fig. \ref{FigS13}, for a uniform strain.

In the context of in-plane tensile expansion ($u_{xx} > 0$ and $u_{yy} > 0$), a consistent trend emerges where the bands are lowered. This trend aligns with the calculations for strain-modified band structures reported in Ref. \cite{Feng2012}. Notably, with a substantial degree of expansion (approximately 0.05), the valence band maxima at the $\Gamma$-point surpasses the energy of the valence band at the K-point, resulting in an indirect bandgap \cite{Dong2014, Wang2021}. A similar realization of an indirect band gap is observed in the scenario involving a $u_{zz}$ strain. In cases of compression, the valence band maxima at $\Gamma$ exceed those at the K-point, whereas in cases of expansion, the conduction band minimum is lower at the Q-point.

For $u_{xy}$, there is not a significant modification in the energies of the bands, and throughout all degree of strain, MoS$_2$ remains a direct gap semiconductor which is in stark contrast to the effects of the other sheer strain components, $u_{xz}$ and $u_{yz}$, which transforms ML-MoS$_2$ into an indirect band gap semiconductor. All in all, strain indeed has an effect of modulating the energy of the electronic bandstructure.

\newpage
\section{Supplementary Information Note 8 \\ Strain dependent hopping integrals}
The extent of orbital overlap plays a crucial role in determining the hopping integral, which is dependent on both the distance between the orbital centers and the orientation of the orbitals $\psi_{li}$ and $\psi_{lj}$:
\begin{align}
t_{lij} &= \braket{\psi_{li}|\hat{H}_l|\psi_{lj}}
\end{align}
While there are methods available for scaling hopping parameters based on orbital distances using the Gr$\ddot{\mathrm{u}}$neisen parameters \cite{PhysRevB.94.155416}, such an approach is not suitable for ML-MoS$_2$ due to its partially three-dimensional bonding nature. The presence of out-of-plane bond angles in ML-MoS$_2$ makes the determination of the hopping matrix sensitive to orientation, having the distance-dependent scaling less applicable. Hence, we resort to using brute-force to calculate the hopping matrix for every strain configuration.

\begin{figure}[h!]
\includegraphics[width = 0.65\textwidth]{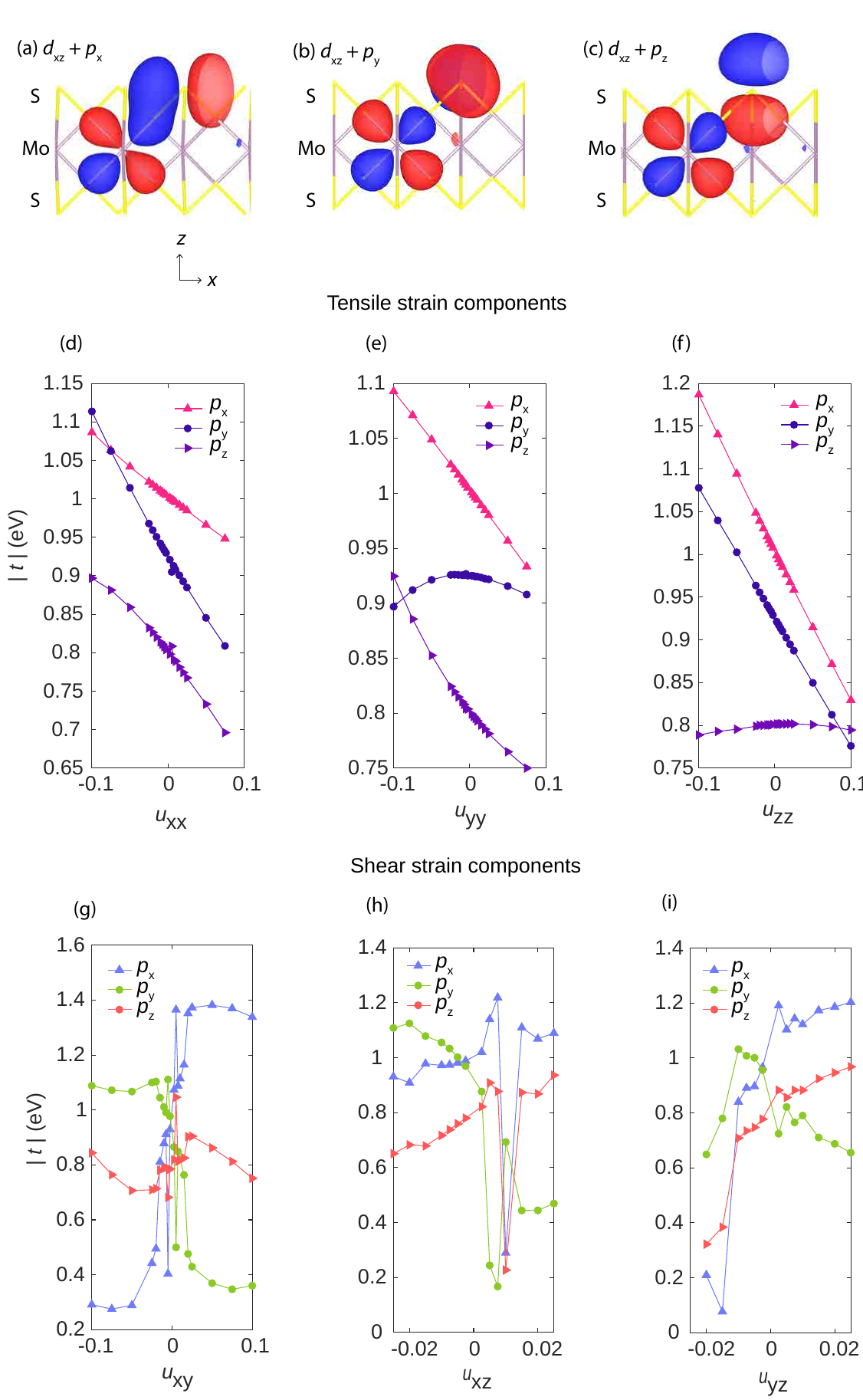}
  \caption { An isosurface plot of the calculated superposition of the Wannier wavefunction basis $\psi =$ (a) $d_{xz} + p_x$, (b) $d_{xz} + p_y$, and (c) $d_{xz} + p_z$, as seen from the $xz$-plane in an unstrained monolayer MoS$_2$. The hopping integral between $d_{xz}$ and $p$ orbitals as a function of (d-f) tensile and (g-i) shear strain components. }
  \label{FigS14}
\end{figure}

To assess the behavior of the calculated change in the hopping parameters due to strain, we focus on a representative NN hopping parameter from the $d_{xz}$ orbital of the Mo atom to orbitals of the S atom (Fig. \ref{FigS14}(a-c)). Generally, under tensile compression ($u_{xx}$, $u_{yy}$, $u_{zz} < 0$), increased overlap yields a higher hopping integral, whereas tensile expansion ($u_{xx}$, $u_{yy}$, $u_{zz} > 0$) lead to reduced overlap and consequently, a decrease in hopping. The strain dependence of the hopping matrix element shown in Fig. \ref{FigS14}(d-f) indeed validates this trend. The change in the hopping matrix element can reach magnitudes on the order of 102 meV under extreme tensile strains of $\pm 0.1$. Notably, hopping from $d_{xy}$ to $p_x$ is generally larger than to $p_y$, followed by $p_z$ (Fig. \ref{FigS14}(d-f)). This is rationalized by the orientation of the orbitals, which allows the $p_x$ orbitals to exhibit stronger overlap with the lobes of the $d_{xz}$ orbitals and the wavefunctions in the adjacent lobes having the same phase, thus allowing for stronger overlap (Fig. \ref{FigS14}(a)).

Furthermore, hopping to py upon applying $u_{yy}$ strain and hopping to $p_z$ upon applying $u_{zz}$ strain results in a non-monotonic strain dependence. This phenomenon arises due to the destructive interference when wavefunctions of opposite phases come closer, leading to a reduction in overlap. Conversely, when these wavefunctions are stretched further apart, those of the same phase overlap less. In contrast to the case of tensile strain (Fig. \ref{FigS14}(d-f), hopping matrix dependence on shear is less systematic, as shear strain induces a change in the orientation of the orbitals, which is dependent on the symmetry of the overlapping orbitals. With extreme sheer of $\pm 0.02$, the hopping can vary by 1 eV (Fig. \ref{FigS14}(d-f)).

\newpage

\newpage
\section{Supplementary Information Note 9 \\ Bias dependence of the LDOS}

\begin{figure}[h!]
\includegraphics[width = 0.5\textwidth]{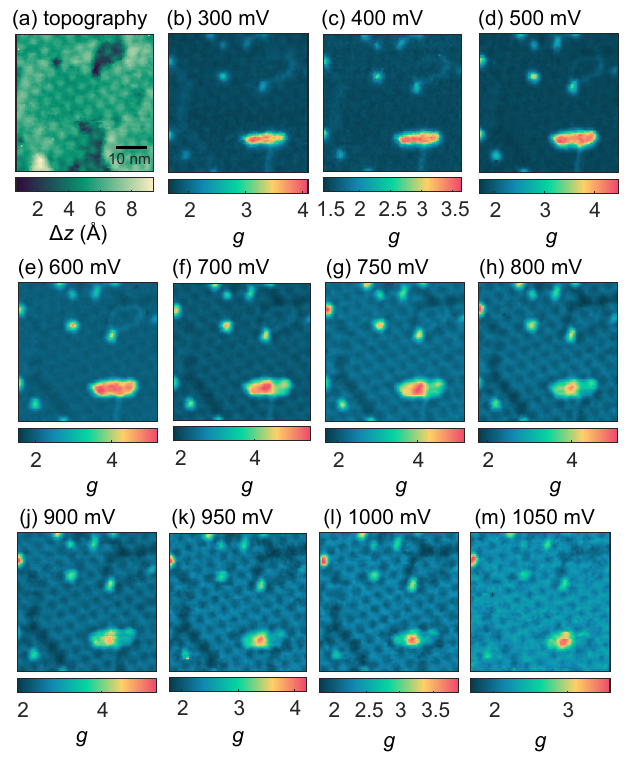}
  \caption { Normalized conductance (g) maps above the Fermi level without the illumination of light: (a) The corresponding topographic image and the constant bias conductance maps at (b) 300 mV, (c) 400 mV, (d) 500 mV, (e) 600 mV, (f) 700 mV, (g) 750 mV, (h) 800 mV, (j) 900 mV, (k) 950 mV, (l) 1000 mV, and (m) 1050 mV. All conductance maps were measured with a tunneling resistance of 1 G$\Omega$. The panels share the same scale bar. }
  \label{FigS15}
\end{figure}

\begin{figure}[h!]
\includegraphics[width = 0.5\textwidth]{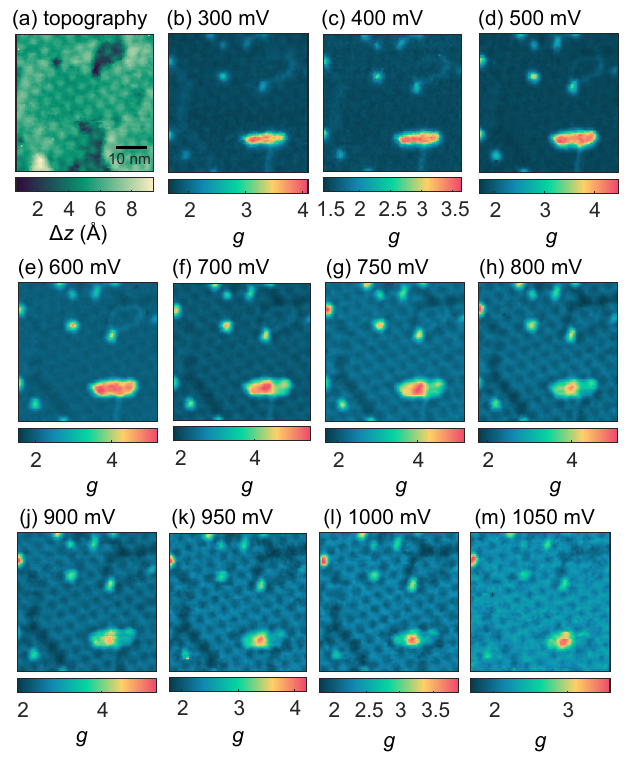}
  \caption { Normalized conductance (g) map below the Fermi level without the illumination of light: (a) The corresponding topographic image and the constant bias conductance map at (b) -500 mV, (c) -550 mV, (d) -600 mV, (e) -650 mV, (f) -700 mV, (g) -750 mV, (h) -800 mV, (j) -850 mV, (k) -900 mV, (l) -950 mV, and (m) -1000 mV. All conductance maps were measured with a tunneling resistance of 1 G$\Omega$. The panels share the same scale bar. }
  \label{FigS16}
\end{figure}

The bias-dependence of the LDOS are shown in Figs. \ref{FigS15} and \ref{FigS16} both taken without the illumination of light. For states above the Fermi level $E_\mathrm{F}$, Moir\'e patterns are visible through the bias range of 300 mV up to 1050 mV (Fig.  \ref{FigS15}). As for states below the valence band, Moir\'e patterns are visible at higher biases (closer to $E_\mathrm{F}$), and become less noticeable well below $E_\mathrm{F}$ (Fig. \ref{FigS16}). In both cases above and below the conduction band, the regions at the edge of corrugations indicated in the topographic images exhibit enhanced LDOS as shown in the conductance map.

\newpage

\newpage
\section{Supplementary Information Note 10 \\ Effect of 1.95 eV laser excitation on ML-MoS$_2$ on Au(111)}
\begin{figure}[h!]
\includegraphics[width = 0.4\textwidth]{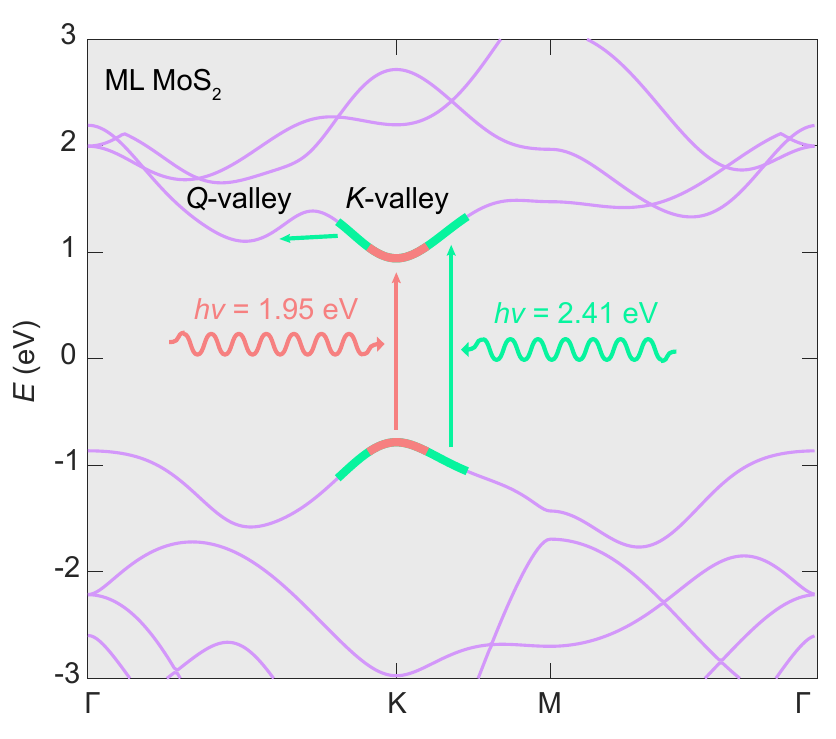}
  \caption {The bandstructure of free-standing ML-MoS$_2$ obtained from MLWFs: The highlighted regions of the bands indicate regions where $E_\mathrm{g}$ is less than $E_\mathrm{photon}$, for the case of 1.95 eV (red) and 2.41 eV (green) photons.}
  \label{FigS17}
\end{figure}

By illuminating the sample with 635 nm laser ($E_\mathrm{photon}$ = 1.95 eV), photoexcited electrons are no longer populated to the Q-valley, but only populate the K-valley among flat free- standing monolayer MoS$_2$ as depicted in Fig. \ref{FigS17}. Even for strained monolayer MoS$_2$, the photoexcited carriers are only limited to lower energy bands by illumination of 635 nm laser, limiting their region of diffusion through the modulated band landscape.

\begin{figure}[h!]
\includegraphics[width = \textwidth]{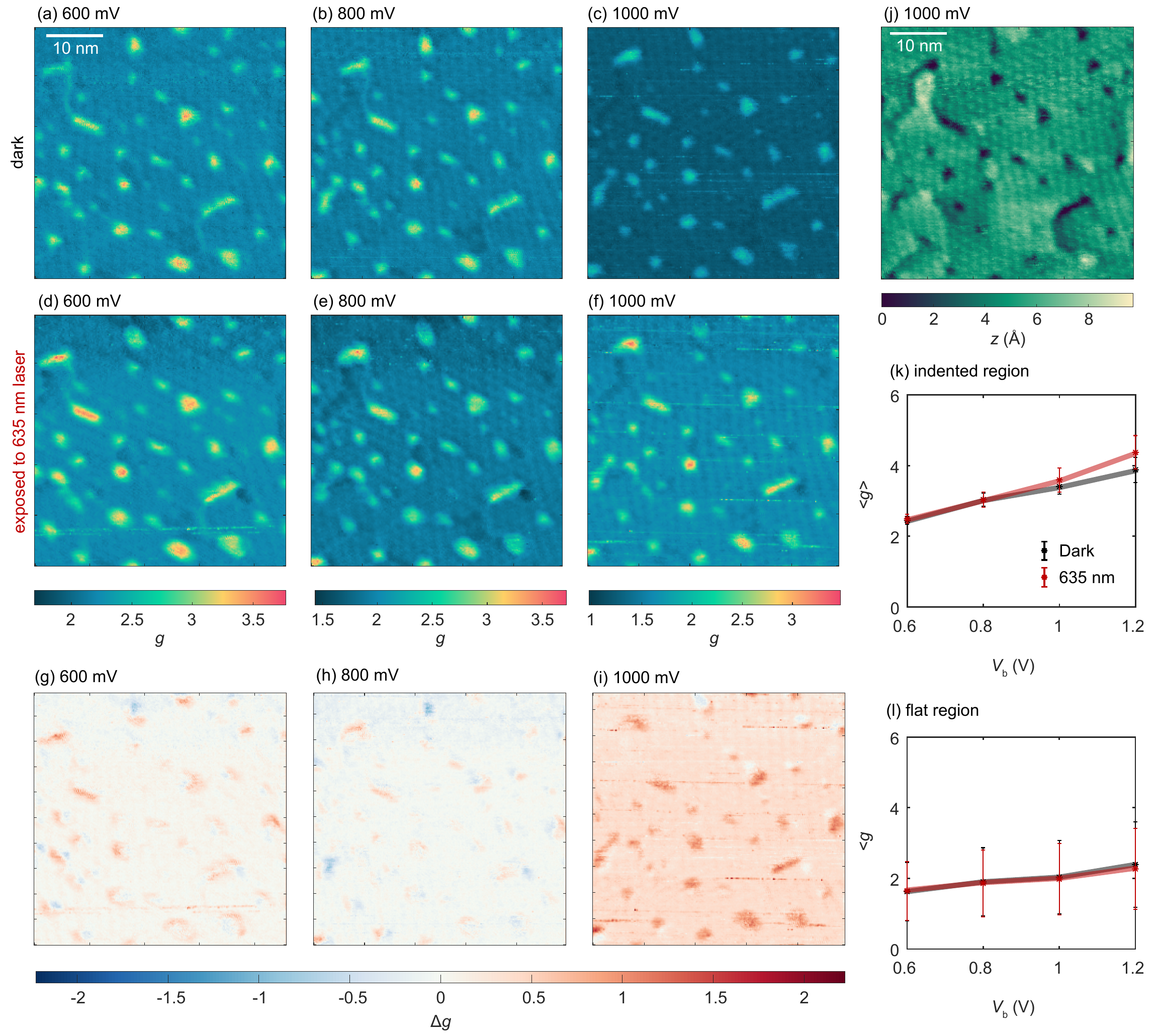}
  \caption {Constant bias map $g(\vec{r}, V_\mathrm{b})$ obtained at various bias voltage (a-c) without the illumination of light, and (d-f) under the illumination of 635 nm light (with 7 mW/cm$^2$ power), measured while maintaining a tunneling resistance of 1 G$\Omega$. A line noise artifact removal was performed to enhance contrast. (g-i) The difference in the conductance map, $\Delta g$, taken with and without the illumination of light at different biases. (j) Topographic map of the areas corresponding to the conductance maps measured with a set point of $V_\mathrm{b}$ = 800 mV and $I_\mathrm{t}$ = 800 pA. (k) The bias dependence on the average g at the indented region, and (l) at the flat region. The error bars indicate the first standard deviation, and a best fit line is drawn to depict the trend. Here we note that the light-induced renormalization effect is largely absent in the flat region and is only noticeable at higher bias voltages in the indented region.}
  \label{FigS18}
\end{figure}

Figure \ref{FigS18} displays constant bias conductance maps $g(\vec{r}, V_\mathrm{b})$ with and without the illumination of $\lambda = 635$ nm laser. As evidenced from Fig. \ref{FigS18}(l,k), there is no significant light-induced changes in the LDOS of ML-MoS$_2$ under $\lambda = 635$ nm laser, which is in sharp contrast to the findings upon the illumination of $\lambda = 515$ nm laser. However, there are still noticeable although slight changes in the LDOS, particularly at higher biases, among the indented regions of the sample.

\newpage

\putbib
\defaultbibliographystyle{apsrev4-2}
\defaultbibliography{apssamp}
\end{bibunit}

\end{document}